\documentclass[10pt, journal, compsoc]{IEEEtran}

\pdfoutput=1
\usepackage{graphicx}
\usepackage{dblfloatfix}
\usepackage[breaklinks=false, allcolors=black, colorlinks=true]{hyperref}

\usepackage{amsfonts}
\usepackage{amsmath} 
\usepackage{amssymb}
\usepackage{amsthm}	
\usepackage{enumerate}
\usepackage{mathrsfs}
\usepackage{caption}
\usepackage{subcaption}
\usepackage{bm}
\usepackage{mathtools}

\theoremstyle{definition} 
\newtheorem{theore}{Theorem}
\newtheorem{corollar}{Corollary}
\newtheorem{lemm}[theore]{Lemma}
\newtheorem{definitio}{Definition}
\newtheorem{proble}{Problem}
\newtheorem{assumptio}{Assumption}
\newtheorem{propert}{Property}
\newtheorem{propositio}{Proposition}
\newtheorem{conjectur}{Conjecture}
\newtheorem{exampl}{Example}
\newtheorem{rul}{Rule}

\newtheorem{remar}{Remark}
\newtheorem{fac}{Fact}

\def\nn{\nonumber}
\def\beq{\begin{equation}}
\def\eeq{\end{equation}}
\def\beqs{\begin{equation*}}
\def\eeqs{\end{equation*}}
\def\balign{\begin{align}}
\def\ealign{\end{align}}
\def\bea{\begin{eqnarray}}
\def\eea{\end{eqnarray}}
\def\ba{\begin{array}}
\def\ea{\end{array}}
\def\bcenter{\begin{center}}
\def\ecenter{\end{center}}
\def\bitem{\begin{itemize}}
\def\eitem{\end{itemize}}
\def\benum{\begin{enumerate}}
\def\eenum{\end{enumerate}}
\def\bdoc{\begin{document}}
\def\edoc{\end{document}}
\def\defeq{{\stackrel{\Delta}{=}}}
\def\nn{\nonumber}

\def\eg{{e.g., \/}}
\def\etal{{\it et al. \/}}
\def\viz{{\iet viz.,\ \/}}
\def\ie{{i.e.,\ \/}}
\def\etc{{\it etc.,\ \/}}
\def\cf{{cf.,\ \/}}

\usepackage[usenames,dvipsnames,svgnames,table]{xcolor}

\def\black{\textcolor{black}}
\def\yellow{\textcolor{yellow}}
\def\green{\textcolor{green}}
\def\tcbg{\textcolor{bgrd}}
\def\magenta{\textcolor{magenta}}
\def\white{\textcolor{white}}
\def\gray{\textcolor{gray}}
\def\cyan{\textcolor{cyan}}

\newcommand\Quote[1]{\lq\textsl{#1}\rq}
\newcommand\fr[2]{{\textstyle\frac{#1}{#2}}}

\newcommand{\Ambb}{\mathbb{A}}
\newcommand{\Bmbb}{\mathbb{B}}
\newcommand{\Cmbb}{\mathbb{C}}
\newcommand{\Dmbb}{\mathbb{D}}
\newcommand{\Embb}{\mathbb{E}}
\newcommand{\Fmbb}{\mathbb{F}}
\newcommand{\Gmbb}{\mathbb{G}}
\newcommand{\Hmbb}{\mathbb{H}}
\newcommand{\Imbb}{\mathbb{I}}
\newcommand{\Jmbb}{\mathbb{J}}
\newcommand{\Kmbb}{\mathbb{K}}
\newcommand{\Lmbb}{\mathbb{L}}
\newcommand{\Mmbb}{\mathbb{M}}
\newcommand{\Nmbb}{\mathbb{N}}
\newcommand{\Ombb}{\mathbb{O}}
\newcommand{\Pmbb}{\mathbb{P}}
\newcommand{\Qmbb}{\mathbb{Q}}
\newcommand{\Rmbb}{\mathbb{R}}
\newcommand{\Smbb}{\mathbb{S}}
\newcommand{\Tmbb}{\mathbb{T}}
\newcommand{\Umbb}{\mathbb{U}}
\newcommand{\Vmbb}{\mathbb{V}}
\newcommand{\Wmbb}{\mathbb{W}}
\newcommand{\Xmbb}{\mathbb{X}}
\newcommand{\Ymbb}{\mathbb{Y}}
\newcommand{\Zmbb}{\mathbb{Z}}

\newcommand{\Amsc}{\mathcal{A}}
\newcommand{\Bmsc}{\mathcal{B}}
\newcommand{\Cmsc}{\mathcal{C}}
\newcommand{\Dmsc}{\mathcal{D}}
\newcommand{\Emsc}{\mathcal{E}}
\newcommand{\Fmsc}{\mathcal{F}}
\newcommand{\Gmsc}{\mathcal{G}}
\newcommand{\Hmsc}{\mathcal{H}}
\newcommand{\Imsc}{\mathcal{I}}
\newcommand{\Jmsc}{\mathcal{J}}
\newcommand{\Kmsc}{\mathcal{K}}
\newcommand{\Lmsc}{\mathcal{L}}
\newcommand{\Mmsc}{\mathcal{M}}
\newcommand{\Nmsc}{\mathcal{N}}
\newcommand{\Omsc}{\mathcal{O}}
\newcommand{\Pmsc}{\mathcal{P}}
\newcommand{\Qmsc}{\mathcal{Q}}
\newcommand{\Rmsc}{\mathcal{R}}
\newcommand{\Smsc}{\mathcal{S}}
\newcommand{\Tmsc}{\mathcal{T}}
\newcommand{\Umsc}{\mathcal{U}}
\newcommand{\Vmsc}{\mathcal{V}}
\newcommand{\Wmsc}{\mathcal{W}}
\newcommand{\Xmsc}{\mathcal{X}}
\newcommand{\Ymsc}{\mathcal{Y}}
\newcommand{\Zmsc}{\mathcal{Z}}

\newcommand{\Ascr}{\mathscr{A}}
\newcommand{\Bscr}{\mathscr{B}}
\newcommand{\Cscr}{\mathscr{C}}
\newcommand{\Dscr}{\mathscr{D}}
\newcommand{\Escr}{\mathscr{E}}
\newcommand{\Fscr}{\mathscr{F}}
\newcommand{\Gscr}{\mathscr{G}}
\newcommand{\Hscr}{\mathscr{H}}
\newcommand{\Iscr}{\mathscr{I}}
\newcommand{\Jscr}{\mathscr{J}}
\newcommand{\Kscr}{\mathscr{K}}
\newcommand{\Lscr}{\mathscr{L}}
\newcommand{\Mscr}{\mathscr{M}}
\newcommand{\Nscr}{\mathscr{N}}
\newcommand{\Oscr}{\mathscr{O}}
\newcommand{\Pscr}{\mathscr{P}}
\newcommand{\Qscr}{\mathscr{Q}}
\newcommand{\Rscr}{\mathscr{R}}
\newcommand{\Sscr}{\mathscr{S}}
\newcommand{\Tscr}{\mathscr{T}}
\newcommand{\Uscr}{\mathscr{U}}
\newcommand{\Vscr}{\mathscr{V}}
\newcommand{\Wscr}{\mathscr{W}}
\newcommand{\Xscr}{\mathscr{X}}
\newcommand{\Yscr}{\mathscr{Y}}
\newcommand{\Zscr}{\mathscr{Z}}

\def\Atiny{\mbox{\tiny{A}}}
\def\Btiny{\mbox{\tiny{B}}}
\def\Ctiny{\mbox{\tiny{C}}}
\def\Dtiny{\mbox{\tiny{D}}}
\def\Etiny{\mbox{\tiny{E}}}
\def\Ctiny{\mbox{\tiny{C}}}
\def\Ftiny{\mbox{\tiny{F}}}
\def\Gtiny{\mbox{\tiny{G}}}
\def\Htiny{\mbox{\tiny{H}}}
\def\Itiny{\mbox{\tiny{I}}}
\def\Jtiny{\mbox{\tiny{J}}}
\def\Ktiny{\mbox{\tiny{K}}}
\def\Ltiny{\mbox{\tiny{L}}}
\def\Mtiny{\mbox{\tiny{M}}}
\def\Ntiny{\mbox{\tiny{N}}}
\def\Otiny{\mbox{\tiny{O}}}
\def\Ptiny{\mbox{\tiny{P}}}
\def\Qtiny{\mbox{\tiny{Q}}}
\def\Rtiny{\mbox{\tiny{R}}}
\def\Stiny{\mbox{\tiny{S}}}
\def\Ttiny{\mbox{\tiny{T}}}
\def\Utiny{\mbox{\tiny{U}}}
\def\Vtiny{\mbox{\tiny{V}}}
\def\Wtiny{\mbox{\tiny{W}}}
\def\Xtiny{\mbox{\tiny{X}}}
\def\Ytiny{\mbox{\tiny{Y}}}
\def\Ztiny{\mbox{\tiny{Z}}}

\def\alphabf{\hbox{\boldmath$\alpha$\unboldmath}}
\def\betabf{\hbox{\boldmath$\beta$\unboldmath}}
\def\gammabf{\hbox{\boldmath$\gamma$\unboldmath}}
\def\deltabf{\hbox{\boldmath$\delta$\unboldmath}}
\def\epsilonbf{\hbox{\boldmath$\epsilon$\unboldmath}}
\def\zetabf{\hbox{\boldmath$\zeta$\unboldmath}}
\def\etabf{\hbox{\boldmath$\eta$\unboldmath}}
\def\iotabf{\hbox{\boldmath$\iota$\unboldmath}}
\def\kappabf{\hbox{\boldmath$\kappa$\unboldmath}}
\def\lambdabf{\hbox{\boldmath$\lambda$\unboldmath}}
\def\mubf{\hbox{\boldmath$\mu$\unboldmath}}
\def\nubf{\hbox{\boldmath$\nu$\unboldmath}}
\def\xibf{\hbox{\boldmath$\xi$\unboldmath}}
\def\pibf{\hbox{\boldmath$\pi$\unboldmath}}
\def\rhobf{\hbox{\boldmath$\rho$\unboldmath}}
\def\sigmabf{\hbox{\boldmath$\sigma$\unboldmath}}
\def\taubf{\hbox{\boldmath$\tau$\unboldmath}}
\def\upsilonbf{\hbox{\boldmath$\upsilon$\unboldmath}}
\def\phibf{\hbox{\boldmath$\phi$\unboldmath}}
\def\chibf{\hbox{\boldmath$\chi$\unboldmath}}
\def\psibf{\hbox{\boldmath$\psi$\unboldmath}}
\def\omegabf{\hbox{\boldmath$\omega$\unboldmath}}
\def\Sigmabf{\hbox{$\bf \Sigma$}}
\def\Upsilonbf{\hbox{$\bf \Upsilon$}}
\def\Omegabf{\hbox{$\bf \Omega$}}
\def\Deltabf{\hbox{$\bf \Delta$}}
\def\Gammabf{\hbox{$\bf \Gamma$}}
\def\Thetabf{\hbox{$\bf \Theta$}}
\def\Lambdabf{\mbox{$ \bf \Lambda $}}
\def\Xibf{\hbox{\bf$\Xi$}}
\def\Pibf{{\bf \Pi}}
\def\thetabf{{\mbox{\boldmath$\theta$\unboldmath}}}
\def\Upsilonbf{\hbox{\boldmath$\Upsilon$\unboldmath}}
\newcommand{\Phibf}{\mbox{${\bf \Phi}$}}
\newcommand{\Psibf}{\mbox{${\bf \Psi}$}}

\def\abf{{\bf a}}
\def\bbf{{\bf b}}
\def\cbf{{\bf c}}
\def\dbf{{\bf d}}
\def\ebf{{\bf e}}
\def\fbf{{\bf f}}
\def\gbf{{\bf g}}
\def\hbf{{\bf h}}
\def\ibf{{\bf i}}
\def\jbf{{\bf j}}
\def\kbf{{\bf k}}
\def\lbf{{\bf l}}
\def\mbf{{\bf m}}
\def\nbf{{\bf n}}
\def\obf{{\bf o}}
\def\pbf{{\bf p}}
\def\qbf{{\bf q}}
\def\rbf{{\bf r}}
\def\sbf{{\bf s}}
\def\tbf{{\bf t}}
\def\ubf{{\bf u}}
\def\vbf{{\bf v}}
\def\wbf{{\bf w}}
\def\xbf{{\bf x}}
\def\ybf{{\bf y}}
\def\zbf{{\bf z}}
\def\rbf{{\bf r}}
\def\xbf{{\bf x}}
\def\ybf{{\bf y}}
\def\Abf{{\bf A}}
\def\Bbf{{\bf B}}
\def\Cbf{{\bf C}}
\def\Dbf{{\bf D}}
\def\Ebf{{\bf E}}
\def\Fbf{{\bf F}}
\def\Gbf{{\bf G}}
\def\Hbf{{\bf H}}
\def\Ibf{{\bf I}}
\def\Jbf{{\bf J}}
\def\Kbf{{\bf K}}
\def\Lbf{{\bf L}}
\def\Mbf{{\bf M}}
\def\Nbf{{\bf N}}
\def\Obf{{\bf O}}
\def\Pbf{{\bf P}}
\def\Qbf{{\bf Q}}
\def\Rbf{{\bf R}}
\def\Sbf{{\bf S}}
\def\Tbf{{\bf T}}
\def\Ubf{{\bf U}}
\def\Vbf{{\bf V}}
\def\Wbf{{\bf W}}
\def\Xbf{{\bf X}}
\def\Ybf{{\bf Y}}
\def\Zbf{{\bf Z}}
\def\Ac{{\cal A}}
\def\Bc{{\cal B}}
\def\Cc{{\cal C}}
\def\Dc{{\cal D}}
\def\Ec{{\cal E}}
\def\Fc{{\cal F}}
\def\Gc{{\cal G}}
\def\Hc{{\cal H}}
\def\Ic{{\cal I}}
\def\Jc{{\cal J}}
\def\Kc{{\cal K}}
\def\Lc{{\cal L}}
\def\Mc{{\cal M}}
\def\Nc{{\cal N}}
\def\Oc{{\cal O}}
\def\Pc{{\cal P}}
\def\Qc{{\cal Q}}
\def\Rc{{\cal R}}
\def\Sc{{\cal S}}
\def\Tc{{\cal T}}
\def\Uc{{\cal U}}
\def\Vc{{\cal V}}
\def\Wc{{\cal W}}
\def\Xc{{\cal X}}
\def\Yc{{\cal Y}}
\def\Zc{{\cal Z}}

\def\0bf{\mathbf{0}}
\def\1bf{\mathbf{1}}

\def\lambdabf{\boldsymbol{\lambda}}
\def\mubf{\boldsymbol{\mu}}
\def\gammabf{\boldsymbol{\gamma}}
\def\xibf{\boldsymbol{\xi}}
\def\sigmabf{\boldsymbol{\sigma}}
\def\zetabf{\boldsymbol{\zeta}}

\def\Lambdabf{\boldsymbol{\Lambda}}
\def\Mubf{\boldsymbol{\Mu}}
\def\Gammabf{\boldsymbol{\Gamma}}
\def\Xibf{\boldsymbol{\Xi}}
\def\Sigmabf{\boldsymbol{\Sigma}}
\def\Zetabf{\boldsymbol{\Zeta}}

\newcommand{\bsf}[1]{\textsf{\textbf{#1}}}
\newcommand{\lbsf}[1]{\textsf{\large  \textbf{#1}}}
\newcommand{\Lbsf}[1]{\textsf{\Large  \textbf{#1}}}
\newcommand{\hbsf}[1]{\textsf{\huge  \textbf{#1}}}

\newcommand{\myminipage}[3]{\begin{minipage}[#1]{#2}{#3} \end{minipage}}
\newcommand{\sbs}[4]{\myminipage{c}{#1}{#3} \hfill
\myminipage{c}{#2}{#4}}

\newcommand{\myfig}[2]{\centerline{\psfig{figure=#1,width=#2,silent=}}}
\newcommand{\myfigh}[2]{\centerline{\psfig{figure=#1,height=#2,silent=}}}
\newcommand{\myfigwh}[3]{\centerline{\psfig{figure=#1,width=#2,height=#3,silent=}}}

\newcommand{\beqa}{\begin{eqnarray}}
\newcommand{\eeqa}{\end{eqnarray}}
\newcommand{\beqan}{\begin{eqnarray*}}
\newcommand{\eeqan}{\end{eqnarray*}}
\newcommand{\dst}[1]{\displaystyle{ #1 }}
\newcounter{pf}

\newcommand{\smax}[1] { \bar \sigma \left( #1 \right) }
\newcommand{\Rn}{{\mathbb R}^n}
\newcommand{\R}{{\mathbb R}}
\newcommand{\C}{{\mathbb C}}
\newcommand{\Rm}{\mathbb{R}^m}
\newcommand{\Rmn}{\mathbb{R}^{m \times n}}
\newcommand{\Rpq}{\mathbb{R}^{p \times q}}
\newcommand{\Cn}{\mathbb{C}^n}
\newcommand{\Cm}{\mathbb{C}^m}
\newcommand{\Cnn}{\mathbb{C}^{n \times n}}
\newcommand{\Cmn}{\mathbb{C}^{m \times n}}
\newcommand{\ip}[1]{\left\langle #1 \right\rangle}
\newcommand{\rank}{\mbox{rank}}
\newcommand{\Span}{\mbox{\rm Span }}
\newcommand{\Trace}{\mbox{\rm Tr }}
\newcommand{\trace}[1]{\text{Tr}\left(#1\right)}
\newcommand{\Spec}{\mbox{\rm Spec }}
\newcommand{\vectornorm}[1]{\left\|#1\right\|}

\newcommand{\pd}[2]{\frac{\partial #1}{\partial #2}}
\newcommand{\ppd}[3]{\frac{\partial^2 #1}{\partial #2 \partial #3}}

\newcommand{\thtilde}{\tilde{\theta}}
\newcommand{\thnom}{\theta^\circ}
\newcommand{\thopt}{\theta^{\mbox{\small opt}}}
\newcommand{\thhat}{{\hat{\theta}}}
\newcommand{\Tho}{\Theta^\circ}
\newcommand{\tho}{\theta^\circ}
\newcommand{\np}{{n_p}}

\newcommand{\ii}{{[i]}}
\newcommand{\II}{{[i+1]}}
\newcommand{\iii}{{[ii]}}
\newcommand{\jj}{{[j]}}
\newcommand{\kk}{{[k]}}
\newcommand{\thi}{{\theta^\ii}}
\newcommand{\thI}{{\theta^\II}}
\newcommand{\di}{{d^\ii}}
\newcommand{\gi}{{g^\ii}}
\newcommand{\Hi}{{\HH^\ii}}
\newcommand{\thK}{\theta^{(k+1)}}
\newcommand{\gk}{{g^{(k)}}}
\newcommand{\Hk}{{{\cal H}^{(k)}}}

\newcommand{\bfdelta}{{\bf \Delta}}

\newcommand{\Exp}[1]{\exp \left\{ #1 \right\}} 
\newcommand{\gaussian}[1]{\mathbb{N} \left( #1 \right)}
\newcommand{\uniform}[1]{\mathbb{U} \left[ #1 \right]}
\newcommand{\exponential}[1]{\mathbb{E} \left[ #1 \right]}
\newcommand{\EXP}[1]{\EEXP \left[ #1 \right]} 
\newcommand{\EEXP}{\mbox{\bsf{E}}} 
\newcommand{\Prob}[1]{\mbox{{\sf Pr}} \left(#1 \right)}
\newcommand{\convas}{\stackrel{as}{\longrightarrow}}
\newcommand{\convinp}{\stackrel{p}{\longrightarrow}}
\newcommand{\convind}{\stackrel{d}{\longrightarrow}}
\newcommand{\convqm}{\stackrel{qm}{\longrightarrow}}
\newcommand{\sss}[1]{{_{#1}}}
\newcommand{\density}[2]{p_{_{_{#1}}}\!\!\left(#2 \right)} 
\newcommand{\distro}[2]{P_{_{_{#1}}}\!\!\left(#2 \right)} 
\newcommand{\rxx}[1]{R_{_{#1}}\!} 
\newcommand{\sxx}[1]{S_{_{#1}}} 
\newcommand{\cov}[1]{\Lambda_{_{#1}}} 
\newcommand{\mean}[1]{m_{_{#1}}} 
\newcommand{\LS}[1]{\hat{#1}_{_{LS}}} 
\newcommand{\MV}[1]{\hat{#1}_{_{MV}}} 
\newcommand{\LMV}[1]{\hat{#1}_{_{LMV}}} 
\newcommand{\ML}[1]{\hat{#1}_{_{ML}}} 

\renewcommand{\arraystretch}{0.9}
\newcommand{\bmat}[1]{ \begin{bmatrix} #1 \end{bmatrix}}
\newcommand{\mat}[1]{ \left[ \begin{array}{cccccccc} #1 \end{array}
\right] }
\newcommand{\smallmat}[1]{\small{\mat{#1}}}
\newcommand{\sysblk}[4]{\begin{array}{c|cccc}#1&#2\\ \hline#3&#4
\end{array}}
\newcommand{\sysmat}[4]{\left[\sysblk{#1}{#2}{#3}{#4}\right]}
\newcommand{\SGeq}{\succ}
\newcommand{\SLeq}{\prec}
\newcommand{\Geq}{\succeq}
\newcommand{\Leq}{\preceq}

\newcommand{\Bset}{\mathbb{B}}
\newcommand{\Cset}{\mathbb{C}}
\newcommand{\Fset}{\mathbb{F}}
\newcommand{\Mset}{\mathbb{M}}
\newcommand{\Nset}{\mathbb{N}}
\newcommand{\Qset}{\mathbb{Q}}
\newcommand{\Rset}{\mathbb{R}}
\newcommand{\Sset}{\mathbb{S}}
\newcommand{\Tset}{\mathbb{T}}
\newcommand{\Uset}{\mathbb{U}}
\newcommand{\Vset}{\mathbb{V}}
\newcommand{\Wset}{\mathbb{W}}
\newcommand{\Zset}{\mathbb{Z}}

\newcommand{\Ical}{{\cal I}}
\newcommand{\Acal}{{\cal A}}
\newcommand{\Bcal}{{\cal B}}
\newcommand{\Ccal}{{\cal C}}
\newcommand{\Dcal}{{\cal D}}
\newcommand{\Ecal}{{\cal E}}
\newcommand{\Fcal}{{\cal F}}
\newcommand{\Gcal}{{\cal G}}
\newcommand{\Hcal}{{\cal H}}
\newcommand{\Kcal}{{\cal K}}
\newcommand{\Lcal}{{\cal L}}
\newcommand{\Mcal}{{\cal M}}
\newcommand{\Ncal}{{\cal N}}
\newcommand{\Pcal}{{\cal P}}
\newcommand{\Qcal}{{\cal Q}}
\newcommand{\Rcal}{{\cal R}}
\newcommand{\Scal}{{\cal S}}
\newcommand{\Tcal}{{\cal T}}
\newcommand{\Wcal}{{\cal W}}
\newcommand{\Ucal}{{\cal U}}
\newcommand{\Vcal}{{\cal V}}
\newcommand{\Xcal}{{\cal X}}
\newcommand{\Zcal}{{\cal Z}}

\newcommand{\EE}{{\bf E}}
\newcommand{\FF}{{\bf F}}
\newcommand{\GG}{{\bf G}}
\newcommand{\HH}{{\bf H}}
\newcommand{\LL}{{\bf L}}
\newcommand{\NN}{{\bf N}}
\newcommand{\MM}{{\bf M}}
\newcommand{\PP}{{\bf P}}
\newcommand{\QQ}{{\bf Q}}
\newcommand{\RR}{{\bf R}}
\renewcommand{\SS}{{\bf S}}
\newcommand{\TT}{{\bf T}}
\newcommand{\VV}{{\bf V}}
\newcommand{\WW}{{\bf W}}

\newcommand{\thk}{\theta^{(k)}}
\newcommand{\thb}{\theta^{\rm opt}}
\newcommand{\alb}{\alpha^{\rm opt}}
\newcommand{\dk}{d^{(k)}}
\newcommand{\Hinf}{{\cal H}_\infty}
\newcommand{\Htwo}{{\cal H}_2}

\renewcommand{\arraystretch}{1.1}

\newcommand{\red}[1]{{\color{red} #1}}
\newcommand{\blue}[1]{{\color{Blue} #1}}

\newcounter{l1}
\newcounter{l2}
\newcounter{l3}
\newcommand{\bdotlist}{\begin{list}{$\bullet$}{}}
\newcommand{\bboxlist}{\begin{list}{$\Box$}{}}
\newcommand{\bbboxlist}{\begin{list}{\raisebox{.005in}{{\tiny
$\blacksquare$ \ \ }}}{}}
\newcommand{\bdashlist}{\begin{list}{$-$}{} }
\newcommand{\blist}{\begin{list}{}{} }
\newcommand{\barablist}{\begin{list}{\arabic{l1}}{\usecounter{l1}}}
\newcommand{\balphlist}{\begin{list}{(\alph{l2})}{\usecounter{l2}}}
\newcommand{\bAlphlist}{\begin{list}{\Alph{l2}.}{\usecounter{l2}}}
\newcommand{\bdiamlist}{\begin{list}{$\diamond$}{}}
\newcommand{\bromalist}{\begin{list}{(\roman{l3})}{\usecounter{l3}}}

\newcommand{\thm}[1]{\noindent \begin{theorem} #1   \end{theorem}}
\newcommand{\prop}[1]{\begin{proposition} #1 \end{proposition}}
\newcommand{\lem}[1]{\begin{lemma} #1  \hfill $\blacksquare$ \end{lemma}}
\newcommand{\ex}[1]{\begin{example} {\rm #1} \end{example}}
\newcommand{\prf}[1]{ \noindent {\em Proof:} \, #1 \hfill $\blacksquare$}
\newcommand{\rem}[1]{\begin{remark} {\rm #1} \hfill $\Box$ \end{remark}}
\newcommand{\defn}[1]{\begin{definition} {\rm #1 } \end{definition}}
\newcommand{\prob}[1]{\begin{exercise} {\rm  #1 } \end{exercise}}
\newcommand{\cor}[1]{\begin{corollary}   #1  \end{corollary}}

\newcommand{\argmin}{\mathop{\rm argmin}}
\newcommand{\argmax}{\mathop{\rm argmax}}
\newcommand{\diag}{\mathop{\mathrm{diag}}}
\newcommand{\tr}{\mathop{\rm Tr}}
\newcommand{\conv}{\mathop{\rm conv}}
\newcommand{\var}{\mathop{\rm Var}}
\renewcommand{\b}[1]{\ensuremath{\boldsymbol{\mathrm{#1}}}}
\newcommand{\ms}{{\rm MS}}
\newcommand{\tcs}{{\rm TCS}}
\newcommand{\scs}{{\rm SCS}}

\newcommand{\E}[1]{\b{\mu}_{{#1}}}
\newcommand{\Var}[1]{{\Sigma_{#1}}}

\newcommand{\bx}{{\bf x}}
\newcommand{\by}{{\bf y}}
\newcommand{\bd}{{\bf d}}
\newcommand{\bone}{\mathbf{1}}

\newcommand{\poa}{{\rm PoA}}
\newcommand{\nash}{{\rm NE}}
\newcommand{\rsi}{{\rm RSI}}
\newcommand{\lerner}{{\rm LI}}
\newcommand{\cost}{{\rm Cost}}
\newcommand{\sign}{{\rm sgn}}

\def\u{u}	
\def\U{U}	
\def\v{v}	
\def\V{V}	
\def\s{s}	
\def\S{S}	
\def\f{q}	
\def\q{q}	
\def\Q{Q}	
\def\c{c}	
\def\z{z}	
\def\xcap{b}
\def\ccap{\mathbf{\c}}
\def\rentftr{\Phi}
\def\rentfsr{\Sigma}
\def\rentDer{\Delta}
\def\lambdao{\lambda^{o}}

\def \add [#1]{\blue{#1}}
\def \replace [#1]#2{\red{#1} \blue{#2}}

\def \wl [#1]{\magenta{(\textbf{Sam says:} #1)}}
\def \eb [#1]{\red{(\textbf{Eilyan says:} #1)}}

\def \edit{\red}

\newcommand{\rone}[1]{\blue{#1}}
\newcommand{\rtwo}[1]{\cyan{#1}}
\newcommand{\remove}[1]{\red{#1}}

\def\capx{K}

\newcommand{\todo}[1]{ \textcolor{red}{(\textbf{To do:}  #1)}}

\ifCLASSOPTIONcompsoc
  \usepackage[nocompress]{cite}
\else
  \usepackage{cite}
\fi

\graphicspath{{figures/}}

\IEEEoverridecommandlockouts

\begin{document}

\title{A Structural Characterization of \\ Market Power  in  Electric Power Networks}

\author{\vspace{.12in} Weixuan Lin   \qquad Eilyan Bitar 
\thanks{Weixuan Lin and Eilyan Bitar are with the School of Electrical and Computer Engineering, Cornell University, Ithaca, NY, 14853, USA.   Emails: {\tt\small \{wl476,eyb5\}@cornell.edu} }
}

\IEEEtitleabstractindextext{%
\begin{abstract}
We consider a market in which capacity-constrained generators compete in scalar-parameterized supply functions to serve an inelastic demand spread throughout a transmission constrained power network. The market clears according to a locational marginal pricing mechanism, in which the independent system operator (ISO) determines the generators' production quantities to minimize the revealed cost of meeting demand, while ensuring that network transmission and generator capacity constraints are met. Under the stylizing assumption that both the ISO and generators choose their strategies simultaneously, we establish the existence of Nash equilibria for the underlying market, and derive an upper bound on the allocative efficiency loss at Nash equilibrium relative to the socially optimal level. We also characterize an upper bound on the markup of locational marginal prices at Nash equilibrium above their perfectly competitive levels. Of particular relevance to ex ante market power monitoring, these bounds reveal the role of certain market structures---specifically, the \emph{market share} and \emph{residual supply index} of a producer---in predicting the degree to which that producer is able to exercise market power to influence the market outcome to its advantage.
Finally, restricting our attention to the simpler setting of a two-node power network, we provide a characterization of market structures under which a Braess-like paradox occurs due to the exercise of market power---that is to say, we provide a necessary and sufficient condition on market structure under which the strengthening of the network's transmission line capacity results in the (counterintuitive) increase in the total cost of generation at Nash equilibrium.
\end{abstract}


}

\maketitle


\IEEEraisesectionheading{\section{Introduction}\label{sec:introduction}}

We consider an electricity market design in which power producers compete in supply functions to meet an inelastic demand distributed throughout a transmission constrained power network. 
Such markets are vulnerable to manipulation given the large leeway afforded producers  in reporting their supply functions \cite{Klemperer1989}. 
The potential for market manipulation is amplified by the largely inelastic nature of electricity demand, and the presence of hard constraints on transmission and production capacity \cite{borenstein2000electricity, bose2015unifying}.
For example, the strategic withholding of generation capacity by certain producers during the 2000-01 California electricity crisis resulted in over 40 billion US dollars in added costs to consumers and businesses, and the bankruptcy of Pacific Gas and Electric  \cite{joskow2002quantitative, weare2003california}.

In this paper, we analyze  a stylized market model in which power producers are  required to bid  supply functions belonging to a scalar-parameterized family  defined  in the manner of \cite{Johari2011, Xu2014}.
Working within this setting, we elucidate the  role of \emph{market structure} in determining the degree to which price-anticipating  producers can exercise \emph{market power} to influence market allocations and prices at equilibrium.

\emph{Related Work:} \
The formal study of (infinite-dimensional) supply function equilibria dates back to the seminal work of Klemperer and Meyer \cite{Klemperer1989}, which revealed that essentially any production profile  can be supported by a supply function equilibrium in the absence of demand uncertainty. 
There has subsequently emerged a large body of literature investigating the existence, uniqueness, and allocative  (in)efficiency of supply function equilibria given various restrictions on the parametric form of supply functions that producers can bid---see, for example,  \cite{Anderson2008, Baldick2002, Baldick2004, baldick2001capacity, ferrero1997transaction, Green1992, Green1996, holmberg2010supply, Li2015,  Rudkevich1998, Rudkevich2005, Vives2011, xiao2015efficiency}.
Closest to the market model considered in the present paper, Johari and Tsitsiklis \cite{Johari2011} propose a scalar-parameterized supply function bidding mechanism in which $N$ producers compete to meet a known and inelastic demand.  Under the assumption that each producer has sufficient capacity to serve the demand individually, they establish an upper bound on the market's corresponding price of anarchy  given by $1 + 1/(N-2)$.  More recently, Xu et al. \cite{Xu2014} provide an elegant generalization of these results to the setting in which producers have limited production capacities, which are encoded  in the supply functions that they bid.

While the previous supply function models  offer a compelling description of competition in single-node electricity markets, the characterization and analysis of supply function equilibria becomes challenging in the presence of network transmission constraints \cite{Wilson2008, Holmberg2013, Hobbs2000, xu2011bidding, berry1999understanding}. 
For example, it was shown in \cite{Hobbs2000} that the profit maximization problem for each producer amounts to a mathematical program with equilibrium constraints (MPEC), which is, in general, a computationally intractable nonconvex optimization problem.
Additionally, it is well known that, even under the restriction to linear or piecewise-constant supply functions, supply function equilibria may fail to exist in simple two or three-node networks  \cite{tang2013game, Liu2007}. 
In an effort to address such difficulties in analysis, there has emerged another stream of literature that resorts to the so-called networked Cournot model to characterize the strategic interaction between producers in constrained transmission networks. We refer the reader to  \cite{Bose2014, cunningham2002empirical, hobbs2001linear, hobbs2007nash, Metzler2003, Neuhoff2005, stoft1999using, Yao2007, Yao2008, Oren1997, Willems2009} for recent advances.

Building on the aforementioned models of competition,  there are  a number of papers in the literature that  empirically investigate the potential emergence of market power and assess the extent to which it exercised by producers in actual electricity markets (e.g., by estimating price-cost markups)  \cite{day2001divestiture, Baldick2004, Green1992, Green1996, borenstein2000competitive, borenstein1999empirical, sweeting2007market, wolfram1999measuring}. 
With regard to market power monitoring in practice, it is not uncommon for regulating authorities to employ a variety of \emph{structural indices} to both assess the potential for market power ex ante, and to detect the actual exercise of market power ex post---see  Twomey et al. \cite{twomey2006review} for a comprehensive overview.
In particular, the residual supply index (RSI) has empirically proven  to be an  effective predictor of market power as measured by  price-cost markup. For example, empirical analyses of hourly market data carried out by the California Independent System Operator (CAISO)  reveal a  significant negative correlation between hourly RSI and hourly price-cost markup \cite{sheffrin2002predicting}. Beyond empirical evidence, however, there is little theoretical justification  in the literature for the observed effectiveness of the RSI as a predictor of market power. An exception to this claim is the earlier work of Newbery \cite{newbery2009predicting}, which establishes an explicit relationship between a producer's Lerner index (at equilibrium) and its RSI in the setting of a single-node Cournot oligopoly model. A basic limitation of these results, however, is their inapplicability to markets with perfectly inelastic demand.

\emph{Our Contribution:} \ 
In this paper, we 
develop a rigorous equilibrium analysis of  the locational marginal pricing mechanism in a general network setting, where generators are required to report scalar-parameterized supply functions.
Adopting a solution concept in which the ISO  and generators move simultaneously, 
we derive upper bounds on the worst-case efficiency loss and price markups seen at Nash equilibria. Of particular relevance to  the design of methods for market power detection and mitigation, these bounds shed light on the explicit role of (and interplay between) certain  structural indices of market power---specifically, the \emph{market share} and \emph{residual supply index} of  a producer---in predicting the extent to which that producer might exercise market power to influence the market outcome to  its advantage, e.g., increasing price above the competitive level  (cf. Theorem \ref{thm:PoA_bound} and Corollary \ref{cor:Lerner}).  In Section \ref{sec:empirical}, we empirically validate the predictive accuracy of our theoretical upper bound on price markups using historical spot price data from the 1999-2000 Great Britain Electricity Pool. 
In Section \ref{sec:case_study}, we specialize our equilibrium analysis to the  setting of a two-node power network, and provide a characterization of market structures under which a \emph{Braess-like  paradox} occurs due to the exercise of market power. That is to say, we characterize a range of  scenarios in which the strengthening of the network's transmission line capacity  results in the (counterintuitive) increase in the total cost of generation at a Nash equilibrium.

\emph{Organization:} \ In Section \ref{sec:formulation}, we introduce the scalar-parameterized supply function bidding mechanism, and formulate the networked supply function game. 
Section \ref{sec:NE} establishes  the existence of Nash equilibria and the uniqueness of the production profile that they induce; along with providing  an upper bound on the worst-case efficiency loss and price markups incurred at Nash equilibria.
In Section \ref{sec:case_study},  we uncover the occurrence  of a transmission expansion  paradox under a specialization of our model to the setting of a two-node power network. Section \ref{sec:conclusion}  concludes the paper with  a discussion on directions for future research.

\emph{Notation:} \  Let $\Rset$ denote the set of real numbers, and $\Rset_+$ the set of non-negative real numbers. Denote the transpose of a vector $x \in \Rset^n$ by $x^\top$. Let $x_{-i} = (x_1, .., x_{i-1}, x_{i+1}, .., x_n) \in \Rset^{n-1} $ be the vector including all but the $i^\text{th}$ element of $x$. Denote by $\bone$ the vector of all ones. 
Denote by $(\cdot)^+$ the positive part function.
For a univariate function $f: \Rset \to \Rset$ that is both left and right differentiable at $x = \overline{x}$, we denote by  $\partial^- f (\overline{x} ) / \partial x$ and $\partial^+ f(\overline{x}) / \partial x$ the left and right derivatives of $f$ evaluated at $x = \overline{x}$, respectively. Finally, the operator $\circ$ denotes the Hadamard product between matrices of the same dimension.

\section{Model and Formulation} \label{sec:formulation}

\subsection{Supply and Demand Models}
We consider the setting in which  producers compete to supply energy to an inelastic demand spread throughout  a transmission constrained power network. The network is assumed to have a connected topology consisting of $n$ transmission buses (or nodes) connected by $m$ transmission lines (or edges). Let $\Vcal := \{1, \dots, n\}$ denote the set of all nodes. 
In addition, we assume that there are $N_i$ producers located at each node $i \in \Vcal$, and let $N := \sum_{i=1}^n N_i$ denote the total number of producers. 
We specify the nodal position of each producer according to an incidence matrix $A \in \{0,1\}^{n \times N}$ that is defined as
\begin{align}
A_{ij} := \begin{cases}
1, & \text{if producer } j \text{ is located at node }i, \\
0, & \text{otherwise}.
\end{cases} \label{eqn:incidence_mat}
\end{align}
Let $\Ncal_i: = \{j \, | \, A_{ij} = 1\}$ be the set of producers at node $i$. 
We assume that each column of the incidence matrix $A$ has exactly one nonzero entry---that is, each producer is located at exactly one node in the network. The formal treatment  of more general generation ownership structures---in which a producer's generation capacity is allowed to span multiple nodes in the network---is left as a direction for future work.

The demand for energy is assumed to be \emph{perfectly inelastic}. Accordingly, we let $d \in \mathbb{R}^n_+$ denote the demand profile across the network,  where $d_i$ represents the demand for energy at node $i$.  We let $x_j$ be the production quantity of producer $j$, and denote by $C_j (x_j)$ the corresponding cost  incurred by producer $j$ for producing $x_j$ units of energy. We denote by $x := (x_1, \dots, x_{N}) \in \Rset^N$ the production profile. We make the following standard assumption regarding the producers' cost functions.
\begin{assumptio}[Convex Production Costs] \label{ass:cost}
The production cost $C_j(x_j)$  of each producer $j \in \{1, \dots, N\}$  is a convex function that satisfies $C_j (x_j) = 0$ for $x_j \leq 0$, and $C_j (x_j) > 0$ for $x_j > 0$.
\end{assumptio}
We also assume that each producer $j \in \{1, \dots, N\}$ has a  \emph{maximum production capacity} $X_j \geq 0$.
It is important to note that Assumption \ref{ass:cost} implicitly ignores producers' ``start-up costs'',  as the incorporation of nonzero startup costs will, in general, result in the discontinuity of producers' production cost functions at the origin.

\subsection{The Economic Dispatch Problem}
Ultimately, the objective of the independent system operator (ISO) is to choose a production profile that minimizes the \emph{true cost} of serving the demand, while respecting the capacity constraints on transmission and generation facilities. Doing so amounts to solving the so called \emph{economic dispatch} (ED) problem, which is formally defined as
\begin{equation}
\begin{alignedat}{8}
&\underset{x \in \Rset^N}{\text{minimize}}  \quad & &\sum_{j=1}^N  C_j (x_j ) \\
&\text{subject to}  \quad & & A x - d \in \Pcal,  \\
&&& 0 \leq x_j \leq X_j, \ \  j = 1, \dots, N.
\end{alignedat} \label{eq:ED}
\end{equation}
Here, $\Pcal \subseteq  \Rset^n$ represents the feasible set of nodal power injections over the network. Adopting the assumptions on which the so called \emph{DC power flow model} \cite{purchala2005usefulness} is based, one can represent the set $\Pcal$ as a polytope 
\begin{align*}
\Pcal = \left\{  \left. y \in \Rset^n \  \right| \ \bone ^\top y = 0, \  H y \leq c   \right\},
\end{align*}
where $H \in \Rset^{2m\times n}$ denotes the shift-factor matrix, and $c \in \Rset^{2m}$ the corresponding vector of transmission line capacities. We will refer to the constraint $\bone ^\top y = 0$ as the \emph{power balance constraint}, and the constraint $H y \leq c$ as the \emph{transmission capacity constraint}.
Any production profile $x^* = (x_1^*, \dots, x_N^*) \in \Rset^N$ that solves \eqref{eq:ED} is called  \emph{efficient}, and the corresponding aggregate production cost $\sum_{j=1}^N  C_j (x_j^*) $   is referred to as the \emph{efficient cost}.

\subsection{Scalar-parameterized Supply Function Bidding}
In practice, the ISO does not have access to the producers' true cost information. Instead, the producers are asked to report their private information to the ISO in the form of supply functions, which specify the maximum quantity a producer is willing to supply as a function of  price.  In the majority of US electricity markets in operation today, it is customary for the ISO to require that  each producer report a supply function in the form of a non-decreasing step function that is parameterized according to a finite number of price-quantity pairs \cite{holmberg2013supply}.  The characterization of market equilibria that might emerge under this class of  supply functions is analytically intractable, in general---even in the absence of network transmission constraints. The difficulty in analysis derives  in large part from the discontinuity of each producer's residual demand function \cite{anderson2002using, anderson2004nash}. As a result, there is a need to resort to stylized supply function equilibrium models, which appropriately restrict the class of supply functions from which a producer is allowed to choose its bid. In principle, restrictions on the class of supply functions should be chosen in such a manner as to facilitate mathematical analyses, while preserving the main structural determinants of market power and the primary mechanisms by which market power is exercised, e.g., through the ``economic withholding'' of capacity.

With this motivation in mind, we investigate the setting in which producers are allowed to bid \emph{scalar-parameterized supply functions}, and analyze the existence and efficiency of market equilibria that result under price-anticipating  behavior in this setting. In particular, we adopt the approach of  Xu et al. \cite{Xu2014}, and consider a capacitated version of the single-parameter supply function first proposed in \cite{Johari2011}.
Specifically, each producer $j$ reports a scalar parameter $\theta_j \in \Rset_+$ that defines a supply function of the form 
\begin{align}
S_j(p ; \theta_j ) =  X_j - \frac{\theta_j}{p},  \label{eqn:supply_func}
\end{align}
where $S_j(p ; \theta_j)$ denotes the maximum quantity that producer $j$ is willing to supply at any price $p >0$.       
Here, $X_j$ is the \emph{true} production capacity of producer $j$. 
Note that, implicit in this choice of supply function parameterization, is the requirement that each producer offer its full capacity into the power market. That is, we do not allow producers to bid their capacities strategically, as the ``physical withholding'' of capacity is carefully monitored and prohibited by the majority of ISOs in operation today \cite{ercot2017, nyiso2016}.
We denote the \emph{strategy profile} of all producers by the vector of bids $\theta := (\theta_1, \dots, \theta_N) \in \Rset_+^N$.

Given the producers' reported bids $\theta$, the ISO's objective is to choose a production allocation that minimizes the reported aggregate production cost, subject to  the network transmission and production capacity constraints. The \emph{reported cost function} of producer $j$ is defined as the integral of its inverse supply function, which is given by
\begin{align}
\widehat{C}_j (x ;  \theta_j ) := \int_0^x \frac{\theta_j}{X_j - z}dz \; =  \;  \theta_j \log \left( \frac{X_j}{X_j - x} \right) . \label{eqn:surrogate_cost_func}
\end{align}

With these reported costs in hand, the ISO solves the following economic dispatch (ED) problem:
\begin{equation}
\begin{alignedat}{8}
&\underset{x \in \Rset^N}{\text{minimize}}  \quad & &\sum_{j=1}^N  \widehat{C}_j (x_j; \theta_j ) \\
&\text{subject to}  \quad & & A x - d \in \Pcal,  \\
&&&  x_j \leq X_j, \  \ j = 1, \dots, N.
\end{alignedat} \label{opt:mkt_clearing}
\end{equation}
Naturally, the misrepresentation of private cost information by  producers has the potential to induce market allocations---as determined by the solution of problem \eqref{opt:mkt_clearing}---that are highly suboptimal (inefficient) for the original ED problem \eqref{eq:ED}. In this paper, we will attempt to understand the role played by different market and network structures  in determining the extent of such  inefficiency at equilibrium.

\subsection{Attributes of the Supply Function Parameterization}

We make several remarks regarding the  supply function parameterization considered in this paper.
First, notice that each producer's reported cost function \eqref{eqn:surrogate_cost_func} resembles a logarithmic barrier function, which encodes each producer's capacity constraint in  a continuously differentiable fashion. 
As a result, one can omit the production capacity constraint associated with any producer $j$ whose bid satisfies $\theta_j > 0$ from the ED problem \eqref{opt:mkt_clearing} without changing its optimal solution. This substantially simplifies the mathematical analysis of the resulting supply function game. Second, the parametric family of supply functions considered in this paper is expressive enough to 
capture  market outcomes in which producers can exercise market power via the ``economic withholding'' of capacity.
Qualitatively, the inverse supply function  under this choice of parameterization resembles the so-called ``hockey stick'' offer strategy in which a producer offers its last few  units of supply at prices that are well in excess of its true
marginal cost \cite{hurlbut2004protecting}.  In this context, $\theta_j/p$ may be interpreted as the amount of capacity withheld by producer $j$ when the price is $p$.  

We also note  that, in contrast to the supply functions  considered in this paper, the more widely studied  family of \emph{affine} supply functions \cite{Rudkevich1998, Baldick2002, baldick2001capacity, Baldick2004} cannot capture the economic withholding of capacity, as producers' inverse supply functions are necessarily affine under this parameterization.  Moreover, the analysis of affine supply function equilibria is known to be analytically intractable in the presence of production capacity constraints \cite{baldick2001capacity}.

\begin{remar}[Possibility of Negative Supply] 
A practical drawback of the class of supply functions that we consider is that they allow for the possibility of  market allocations (i.e., solutions of the ED problem \eqref{opt:mkt_clearing}) in which a producer has a negative supply allocation. 
We will, however, show that such outcomes are not possible at equilibrium. Namely, we show in Proposition \ref{prop:equilibrium} (in Section \ref{sec:NE}) and Proposition \ref{thm:CE} (in Appendix \ref{sec:CE}) that the production quantity of a producer is guaranteed to be non-negative at Nash equilibria and competitive equilibria, respectively.
It is also worth noting that the results of this paper continue to hold under a modified class of  supply functions given by $S_j (p; \theta_j) = \max \left\{ X_j - \frac{\theta_j}{p}, \ -\epsilon \right\}$,
where $\epsilon > 0$ is an arbitrary positive constant. We forgo this  treatment for ease of exposition.
\end{remar}

\subsection{Nodal Decomposition of   Economic Dispatch} \label{sec:primal-decomp}
In what follows, we develop a primal decomposition of the ED problem \eqref{opt:mkt_clearing}, which reveals an explicit relationship between an individual producer's production quantity and the aggregate production quantity at his node. We do so by introducing an auxiliary variable $q := Ax \in \Rset^n$, which we refer to as the \emph{nodal supply profile}. Here, $q_i = \sum_{j \in \Ncal_i} x_j$ represents the aggregate production quantity at node $i$, and serves as the coupling variable between the (network-wide) ED problem and the nodal ED problems defined in terms of  the local variables $\{x_j | j \in \Ncal_i\}$ at each node $i \in \Vcal$. More formally, the ED problem  \eqref{opt:mkt_clearing} admits an equivalent reformulation as 
\begin{equation}
\begin{alignedat}{8}
&\underset{q \in \Rset^n}{\text{minimize}}  \quad & &\sum_{i=1}^n  G_i (q_i; \theta) \\
&\text{subject to}  \quad & & q - d \in \Pcal, \\
&&& q_i = 0, \qquad &&\text{if } N_i = 0, \\
&&& q_i \leq \sum_{j \in \Ncal_i} X_j, \quad \ &&\text{if } N_i > 0, \ \ i = 1, \dots, n,
\end{alignedat} \label{opt:mkt_clearing_q}
\end{equation}
where $G_i (q_i; \theta)$ denotes the optimal value of the \emph{local ED problem} at node $i$. It  is defined as
\begin{equation}
\begin{split}
\hspace{-.1in} G_i (q_i; \theta) := \min \left\{ \sum_{j \in \Ncal_i} \widehat{C}_j (x_j; \theta_j) \left| \sum_{j \in \Ncal_i} x_j = q_i, \right. \right. \qquad \ &\\
x_j \leq X_j, \ \forall j \in \Ncal_i  \Bigg\}.&
\end{split} \label{opt:mkt_clearing_node_i}
\end{equation}

Given a fixed nodal supply profile $q$, the network-wide ED problem \eqref{opt:mkt_clearing_q} can be separated across nodes as $n$ local ED problems \eqref{opt:mkt_clearing_node_i}. Moreover, the optimal solution  to the local ED problem \eqref{opt:mkt_clearing_node_i} can be expressed in closed-form as an explicit function of the nodal supply profile $q$.  It is not difficult to show that
if $\sum_{j \in \Ncal_i} \theta_j > 0$, then the  optimal solution to \eqref{opt:mkt_clearing_node_i} is unique and is given by
\begin{align}
x_j \left( q_i, \theta \right) = X_j - \frac{\theta_j}{\sum_{k \in \Ncal_i} \theta_k} \cdot  \left( \left( \sum_{k \in \Ncal_i} X_k  \right) - q_i \right) \label{eq:production_1}
\end{align}
for each $j \in \Ncal_i$. 
If, on the other hand, $\sum_{j \in \Ncal_i} \theta_j = 0$, then any feasible solution to the local ED problem \eqref{opt:mkt_clearing_node_i} is optimal. We specify one optimal solution to \eqref{opt:mkt_clearing_node_i} according to
\begin{align}
x_j \left( q_i, \theta \right) = \frac{X_j}{\sum_{k \in \Ncal_i} X_k} \cdot q_i \label{eq:production_2}
\end{align}
for each $j \in \Ncal_i$.  With Eqs. \eqref{eq:production_1}--\eqref{eq:production_2} in hand, one can express the production cost at node $i$ in closed-form as $G_i (q_i; \theta) = \sum_{j \in \Ncal_i} \widehat{C}_j (x_j (q_i, \theta) ; \theta_j ).$

\begin{remar}[Nodal Decomposition of Strategy Profile]
We note that $x_j(q_i, \theta)$ depends on the global strategy profile $\theta$ only through the \emph{local strategy profile} $\{\theta_k | k \in \Ncal_i\}$. This reveals an important insight. Namely, given a fixed nodal supply profile $q$, the explicit interaction between producers decouples across the different nodes in the network. Such insight will play an important role in our game theoretic analysis in Section \ref{sec:NE}.
\end{remar}

\subsection{Networked Supply Function Game}\label{sec:game}
We now present a game theoretic model of supply function competition in  a constrained power network. 
We define the set of players as $\Ncal := \{0, 1, \dots, N\}$, where  0 denotes the ISO, and  $j \in\{1, \dots, N\}$ denotes the $j$th power producer. In practice, the producers and ISO engage in a \emph{sequential-move} game in which the producers simultaneously report their bids, in anticipation of the ISO's determination of production quantities and nodal prices according to the solution of the ED problem \eqref{opt:mkt_clearing}. Formally, this amounts to a multi-leader, single-follower Stackelberg game. However, given the generality of the setting considered in this paper, a general equilibrium analysis of a sequential-move formulation is seemingly out of reach.  
We, therefore, adopt a simplifying assumption, and consider a model of competition that assumes that the producers and ISO \emph{move simultaneously}.\footnote{The model of simultaneous movement  adopted in this paper---in which the ISO's strategic  variables are the nodal supply quantities---is known to manifest in market equilibria that \emph{underpredict}  the intensity of competition in power networks with large transmission capacities, as compared to the more plausible sequential-move formulation \cite{Neuhoff2005, Yao2008}. We refer the reader to \cite[Sec. IV]{lin2016parameterized}, which provides a  detailed numerical comparison of market equilibria  that result under both the sequential and simultaneous-move formulations in a two-node network. Both models are shown to  provide identical predictions if the transmission line is congested under the simultaneous-move formulation. However, if the line capacity is sufficiently large such that the transmission line is guaranteed to never congest,  then the simultaneous-move formulation is shown to predict higher price markups and a greater loss of allocative efficiency at equilibrium than is predicted by the sequential-move formulation. In contrast, an alternative model of simultaneous movement---in which the ISO's strategic variables are the nodal price differences---will result in market equilibria that coincide with those predicted by the sequential-move formulation in networks with sufficiently large transmission capacities,  but will \emph{overpredict} the intensity of competition in  networks with limited transmission capacities. We refer the reader to Yao et al.  \cite{Yao2008} for a  comprehensive discussion on the relative advantages and disadvantages of these competing models of simultaneous movement.}
Specifically, we assume that the  producers choose their supply function bids $\theta$ concurrent with the ISO's determination of the nodal supply profile $q$.  As a result,  the networked supply function game decouples across the nodes in the network, where  the producers at each node, taking the ISO's nodal supply quantity as given, compete only amongst themselves in determining their supply function bids. 
 Such a requirement of simultaneous movement can be interpreted as an assumption of ``bounded rationality'' in which the producers only partially anticipate the impact of their bids on the congestion charges (i.e., the nodal prices differences) that result at equilibrium---see, for example, Metzler et al. \cite{Metzler2003} and Yao et al. \cite{Yao2008}.
 We note that, for reasons of computational and analytical tractability, the  assumption of simultaneous movement is commonly employed in the related literature investigating the use of Cournot models to describe competition in transmission constrained power markets \cite{Bose2014, Metzler2003, Neuhoff2005, Yao2007, Yao2008}.
 
 We proceed with  a formal description of the market participants, their strategy sets, and payoff functions in the networked supply function game.

 \subsubsection{Independent System Operator (ISO)}  The ISO chooses the  production quantities of the individual producers to minimize the reported aggregate cost, while respecting transmission and production capacity constraints. Given the nodal decomposition of the ED problem developed in Section \ref{sec:primal-decomp}, such choice can be reduced to the determination of the nodal supply profile $q \in \Rset^n$---which we define to be the strategy of the ISO. 
Accordingly, we define the \emph{payoff of the ISO} as
\begin{align*}
\pi_0 \left(  q, \theta \right) := - \sum_{i=1}^n  G_i (q_i; \theta),
\end{align*}
where his  feasible strategy set is defined as
\begin{align*}
\Xcal_0 := \left\{ q \in \Rset^n \left| q - d \in \Pcal, \ q_i \leq \sum_{j \in \Ncal_i} X_j , \ \text{if } N_i > 0,  \right. \right. & \quad \\
q_i = 0, \ \text{if } N_i = 0 \Bigg\}&.
\end{align*}

\subsubsection{Producers and Nodal Pricing Mechanism}  Each producer $j \in \Ncal_i$  at node $i \in \Vcal$ must choose a bid parameter $\theta_j \geq 0$, which specifies his supply function. With a slight abuse of notation, we denote the production quantity of producer $j$  by $x_j (q, \theta) := x_j (q_i, \theta )$, where the right-hand side is specified according to Eqs. \eqref{eq:production_1}--\eqref{eq:production_2}.

In this paper, we consider a \emph{nodal pricing mechanism}, i.e.,  a mechanism in which prices are allowed to vary across the different nodes in the network.
In particular, if the bids submitted at node $i$ are such that $\sum_{j \in \Ncal_i} \theta_j > 0$, then the corresponding  price at node $i$ is chosen to  \emph{clear the market} at that node. That is to say, the price at node $i$ is set as the \emph{unique} solution to the equation $\sum_{j \in \Ncal_i} S_j (p; \theta_j) = q_i$. 
If, on the other hand, the bids submitted at node $i$ are such that $\sum_{j \in \Ncal_i} \theta_j = 0$, then it follows  that $S_j(p; \theta_j) = X_j$ for all producers $j \in \Ncal_i$, whatever the price $p$. In this case, we set the price equal to zero. It follows that  the price at each node $i \in \Vcal$ is given by 
\begin{align}
p_i \left(  q, \theta \right) =  \begin{cases} 
\dfrac{ \sum_{j \in \Ncal_i} \theta_j }{\left( \sum_{j \in \Ncal_i} X_j  \right) - q_i},   & \text{if} \ \sum_{j \in \Ncal_i} \theta_j > 0 \\
0, & \text{if} \ \sum_{j \in \Ncal_i} \theta_j = 0,
\end{cases}   \label{eq:price}
\end{align} 
given a nodal supply profile $q$ and bid profile $\theta$. 
\begin{remar}[Locational Marginal Pricing]
When the nodal supply profile $q\in \Rset^n$  is chosen to solve the ED problem \eqref{opt:mkt_clearing_q}, the nodal pricing mechanism \eqref{eq:price} corresponds to the so called \emph{locational marginal pricing} (LMP) mechanism used in many electricity markets that are in operation today. 
In Appendix \ref{sec:CE}, we show that the LMP mechanism ensures the existence of an efficient competitive equilibrium in the presence of transmission capacity constraints.
\end{remar}

With the previous specification of nodal production quantities and prices  in hand, we are now in a position to formally define the \emph{payoff of  each producer} $j \in \Ncal_i$  as
\begin{align}
\pi_j \left( q, \theta \right) := p_i \left( q, \theta \right) x_j \left( q, \theta \right)  - C_j \left( x_j \left( q, \theta \right) \right). \label{eqn:payoff_producer}
\end{align}
Producer $j$'s feasible strategy set is given  by $\Xcal_j := \Rset_+$.

\subsubsection{Solution Concept}   Let  $\Xcal := \prod_{j=0}^N \Xcal_j$ denote the feasible strategy set for all players, and   $\pi := (\pi_0, \pi_1, \dots, \pi_N)$ denote their collection of payoff functions. It follows that the
 triple $(\Ncal, \Xcal, \pi)$ defines a normal-form game, which we shall refer to as the \emph{networked supply function game} for the remainder of this paper.
We describe  stable outcomes of the game $(\Ncal, \Xcal, \pi)$ according to the Nash equilibrium solution concept. We restrict our attention to pure strategy Nash equilibria in this paper, as it is straightforward to show that the networked supply function game does not admit any  non-degenerate mixed strategy Nash equilibria under Assumption \ref{ass:cap} (which ensures  the strict concavity of producers' payoff functions).

\begin{definitio}[Nash Equilibrium] \label{def:NE}
The pair $\left( q, \theta \right) \in \Xcal$ is a \emph{pure strategy Nash equilibrium} (NE) of the game $(\Ncal, \Xcal, \pi)$ if the payoff of the ISO satisfies
$$\pi_0 \left( q, \theta \right) \geq \pi_0 \left( \overline{q}, \theta \right) \ \text{for all}  \ \  \overline{q} \in \Xcal_0,$$ 
 and the payoff of each producer $j \in  \{1, \dots, N\}$ satisfies 
$$\pi_j \left( q, \theta_j, \theta_{-j} \right) \geq \pi_j \left( q, \overline{\theta}_j, \theta_{-j} \right)  \ \text{for all}  \ \ \overline{\theta}_j \in \Xcal_j.$$
We let $\Xcal_{\nash} \subseteq \Xcal$ denote the set of all pure strategy Nash equilibria associated with the game $(\Ncal, \Xcal, \pi)$.
\end{definitio}

In Proposition \ref{prop:equilibrium}, we show that the networked supply function game is guaranteed to admit at least one pure strategy Nash equilibrium if Assumptions \ref{ass:cost}-\ref{ass:cap} are satisfied.
Naturally, the production profile at a Nash equilibrium may differ from the efficient production profile. 
We, therefore,  use the  \emph{price of anarchy} as a measure of the \emph{allocative efficiency loss} at a Nash equilibrium \cite{koutsoupias1999worst}.
\begin{definitio}[Price of Anarchy] \label{def:PoA}
The \emph{price of anarchy} (PoA) associated with the game $(\Ncal, \Xcal, \pi)$ is defined  according to
\begin{align*}
\poa := \sup \left\{ \left. \frac{\sum_{j=1}^N  C_j \left( x_j \left( q, \theta \right) \right)}{\sum_{j=1}^N C_j (x_j^*) } \right| \left( q, \theta \right) \in \Xcal_{\nash} \right\},
\end{align*}
where $x^*$ is the efficient production profile.
\end{definitio}

\section{Nash Equilibrium} \label{sec:NE}

In this section, we characterize the set of Nash equilibria of the networked supply function game. 
Specifically, we characterize the production profile and the nodal supply profile at a Nash equilibrium as the unique optimal solutions to two different convex programs, and provide upper and lower bounds on the nodal prices at a Nash equilibrium in Section \ref{sec:charNE}.
In Sections \ref{sec:effloss}--\ref{sec:markup}, we use this characterization to derive  upper bounds on  the worst-case allocative  efficiency loss and  nodal price markups  at a Nash equilibrium. The upper bounds are explanatory in nature, as they reveal an explicit relationship between a producer's market power and classical structural indices of market power---specifically, the producer's \emph{market share} and its \emph{residual supply index}.  Finally, in Section \ref{sec:empirical}, we empirically evaluate the predictive quality of our theoretical upper bound on price markup using historical spot price data from the 1999-2000 Great Britain Electricity Pool.

\subsection{Structural  Market Power Indices}
In what follows, we provide formal definitions of a producer's  \emph{market share} (MS)  and  \emph{residual supply index} (RSI).
In order to define these market power indices,  we first introduce a (worst-case) measure of peak demand at a node, which we refer to as  the \emph{maximum nodal supply}. More precisely, the maximum nodal supply at node $i \in \Vcal$ is defined as
\begin{align}
q_i^{\max} := \sup \left\{ q_i \left| q \in \Rset_+^n, \ q - d \in \Pcal \right. \right\}.
\end{align}
Clearly, the maximum nodal supply at each node depends on both the network's transmission capacity and  the demand profile. And it holds that $d_i \leq q_i^{\max} \leq \mathbf{1}^\top d$ for each node $i \in \Vcal$.
Using this (conservative) measure of peak  demand at  a node, we define the market share of each producer as follows.
\begin{definitio}[Market Share] \label{def:ms}
The \emph{market share} (MS) of a producer $j \in \Ncal_i$ at node $i \in \Vcal$  is defined as
\begin{align*}
\ms_j := \frac{\min \{X_j, q_i^{\max} \}}{q_i^{\max}}.
\end{align*}
\end{definitio}

Note that $\ms_j \in [0,1]$ for all producers, and that producers with large (relative) market shares are more likely to possess the ability to exercise market power.
Despite its prevelant use among market monitors, the market share index has been criticized as an inadequate measure of market power in unconcentrated markets in which the largest producer has a small market share, but  is close to being \emph{pivotal}  \cite{sheffrin2002predicting, rahimi2003effective}.
For instance, Sheffrin \cite{sheffrin2001critical}  argues that California's deregulated electricity markets were \emph{far} from being competitive during the 2000--2001 crisis, in spite of the fact that no single producer had a market share exceeding 20\%. Sheffrin goes on to claim that, on many occasions,  producers with a market share less than   10\%  were able to influence  the market clearing price to an unwarranted degree.

In part, such critiques of the market share index have served to motivate the design of alternative screening tools for market power. One  commonly used screen---originally developed by the California Independent System Operator (CAISO) \cite{sheffrin2002predicting}---is the \emph{residual supply index}. Essentially, the residual supply index of producer $j$ measures the  extent to which the remaining  aggregate production capacity  in the market (excluding producer $j$) is capable of meeting demand. More precisely, we have the following definition.

\begin{definitio}[Residual Supply Index] \label{def:rsi}
For each node $i \in \Vcal$, the \emph{residual supply index} (RSI) of each producer $j \in \Ncal_i$ is defined according to
\begin{align*}
\rsi_j := \frac{\sum_{k \in \Ncal_i} X_k - X_j}{q_i^{\max}}. 
\end{align*}
\end{definitio}
The residual supply index takes values $\rsi_j \in [0, \infty)$. 
According to this definition, producer $j$ is said to be \emph{pivotal} if $\rsi_j  < 1$. That is, the removal of producer $j$ from node $i$ precludes the remaining producers (at that node) from meeting  the maximum nodal supply at node $i$. Clearly, 
there is potential for  producer $j$ to exercise considerable market power if it is pivotal. 
If, on the other hand, $\rsi_j  \gg 1$, then producer $j$ is far from being pivotal, and will likely have little market power to influence the market clearing price.
In practice, the residual supply index has proved to be effective in predicting the exercise of market power in electricity markets
\cite{bataille2014screening, newbery2005market, swinand2010modeling, sheffrin2002predicting, somani2008agent}.
For example, it was shown in \cite{sheffrin2001critical} that market clearing prices are close to being competitive on average if the RSI of the largest producer is no less than 120\%.

\begin{figure*}[b]
\centering
\hrulefill

\begin{align}
\widetilde{C}_j (x_j; q_i  ) \  :=  \   \left( 1 + \frac{x_j}{\sum_{k \in \Ncal_i, k \neq j} X_k - q_i  } \right) C_j (x_j) - \left(\frac{1}{\sum_{k \in \Ncal_i, k \neq j} X_k - q_i  } \right) \int_{0}^{x_j} C_j (z) \mathrm{d} z
 \label{eq:modified_cost} 
\end{align}

\begin{align}
g_i (z) \  := \ \max_{j \in \Ncal_i} \left\{  \frac{\partial^- \widetilde{C}_j (\widetilde{x}_j; z)}{\partial x_j}  \; \left| \; \widetilde{x} \in \argmin_{ x \in \Rset^N} \Bigg\{   \sum_{k \in \Ncal_i} \widetilde{C}_k (x_k; z) \  \Bigg|   \ 
  \sum_{k \in \Ncal_i} x_k = z    \ \   \text{and} \ \ x_k \leq X_k \  \ \forall \  k \in \Ncal_i \Bigg\} \right\} \right.   \label{eq:def_gi(z)}
\end{align}
\end{figure*}

\begin{remar}[Network Structure]
We note that our definition of residual supply index reflects the potential impact that `network structure'  will have on a producer's market power in a \emph{worst-case} sense. First, each producer's RSI is measured with respect to a surrogate for its `nodal demand'  that corresponds to the maximum power injection that can be feasibly supported by the network at the producer's node.  Second, the `residual production capacity' at  each node is calculated with respect to the aggregate production capacity at that node alone, and disregards the potential contribution of production capacity from other nodes. Although they might appear overly conservative at first glance, the combination of these two approximations in characterizing a producer's residual supply index will play a central role in our derivation of upper bounds on the worst-case  allocative efficiency loss and price markups seen at Nash equilibria in the presence of network constraints. 
\end{remar}

\subsection{Characterizing Nash Equilibrium} \label{sec:charNE}
In Proposition \ref{prop:equilibrium}, we establish the \emph{existence of Nash equilibria} for the networked supply function game.
Additionally, we characterize the nodal supply profile and the production profile that result  at a Nash equilibrium as the unique optimal solutions to two different convex programs, and provide upper and lower bounds on the nodal prices seen at a Nash equilibrium. This characterization will play an integral role in our subsequent derivation of  upper bounds on the worst-case allocative efficiency loss and nodal price markups seen at Nash equilibria.
We first require the following assumption, which limits the `local market power' of each producer.

\begin{assumptio}[No Pivotal Supplier] \label{ass:cap}
The residual supply index of each producer $j \in \{1, \dots, N\}$ satisfies $\rsi_j > 1$.
\end{assumptio}

Essentially, Assumption \ref{ass:cap} amounts to requiring that no producer be \emph{pivotal}.\footnote{An important limitation of Assumption \ref{ass:cap} is that it implicitly requires that there is either no producer or at least two producers at each node in the network.  One possible way of ensuring the satisfaction of Assumption \ref{ass:cap}, in practice, is to formulate the   network model according to a ``reduction'' of the  actual power network, where a node in the reduced network corresponds to a 
connected subnetwork  of  buses that are connected by uncongested transmission lines. One approach to constructing such reduced network models  is to leverage on common prior knowledge of transmission lines that are ``systematically congested" in practice, as this would yield a reasonable approximation of the network according to uncongested subnetworks connected by transmission lines that are ``normally'' binding. We refer the reader to Yao et al. \cite{yao2010hybrid} for a more detailed discussion on the treatment of ``systematically congested" transmission lines in constructing such network reductions.}
The requirement of no pivotal supplier is enforced in electricity markets via the performance of the so-called `pivotal supplier screen'. 
In the United States, for instance, a producer is said to pass the pivotal supplier screen if the \emph{annual peak demand} can be met in the absence of this producer \cite[p. 18]{ferc697}. 
Additionally, the violation of this assumption (i.e., the presence of pivotal suppliers) is known to manifest in large price markups and allocative efficiency loss at equilibrium---see \cite{sheffrin2002predicting, wolak2009assessment, genc2011supply, brandts2014pivotal} for several related theoretical and empirical analyses. Moreover, 
in Appendix \ref{app:ex:unbounded_PoA}, we provide an example of a two-node network, which reveals that, in the absence of Assumption \ref{ass:cap}, the allocative efficiency loss at a Nash equilibrium can be arbitrarily large for the game considered in this paper.

With Assumption \ref{ass:cap} in hand, we state the following result, which establishes the existence of Nash equilibria, provides upper and lower bounds on nodal prices at a Nash equilibrium, and shows that the production profile and the nodal supply profile at a Nash equilibrium are uniquely determined as the solutions of two explicit convex programs.

\begin{propositio}[Existence and Characterization of NE]
Let Assumptions \ref{ass:cost}-\ref{ass:cap} hold. 
\begin{enumerate}[(i)] \setlength{\itemsep}{.1in}
\item The networked supply function game $(\Ncal, \Xcal, \pi)$  admits at least one pure strategy Nash equilibrium.

\item The production profile  $x \left( q, \theta \right) \in \Rset^N$ at a Nash equilibrium $(q, \theta)$ is the unique  optimal solution to the following convex program:
\begin{equation}
\begin{alignedat}{8}
&\underset{x \in \Rset^N}{\text{minimize}} \quad & &\sum_{i=1}^n \sum_{j \in \Ncal_i} \widetilde{C}_j (x_j; q_i  )\\
&\text{subject to} \quad & & Ax - d \in \Pcal, \\
&&& 0 \leq x_j \leq X_j, \quad   j = 1, \dots, N,
\end{alignedat} \label{opt:equilibrium}
\end{equation}
where the \emph{modified  cost functions} $\{\widetilde{C}_j (x_j; q_i  )\}_{j=1}^N$ are defined according to Eq. \eqref{eq:modified_cost}.

\item The price $p_i (q, \theta)$ at each node $i \in \Vcal$ at a Nash equilibrium $(q, \theta)$ satisfies
\begin{align}
p_i (q, \theta) & \in  \left[   \frac{\partial^- \widetilde{C}_j }{  \partial x_j} ,   \  \frac{ \partial^+ \widetilde{C}_j }{ \partial x_j}  \right]  \   \text{if } \ x_j (q, \theta) \in [0, X_j) , \label{eq:NEprice_1}\\
p_i (q, \theta) & \in  \left[ \frac{\partial^- \widetilde{C}_j }{  \partial x_j} , \ \infty \right)  \hspace{1.8em} \text{if } \ x_j (q, \theta) = X_j, \label{eq:NEprice_2}
\end{align}
for each producer $j \in \Ncal_i$, where
\begin{align*}
\frac{\partial^- \widetilde{C}_j }{  \partial x_j}   :=   \frac{\partial^- \widetilde{C}_j (x_j (q, \theta); q_i)}{\partial x_j},     
\\
\frac{\partial^+ \widetilde{C}_j }{  \partial x_j}   :=   \frac{\partial^+ \widetilde{C}_j (x_j (q, \theta); q_i)}{\partial x_j}. 
\end{align*}

\item The nodal supply profile $q \in \Rset^n$  at a Nash equilibrium $(q, \theta)$  is the unique optimal solution to the following convex program:
\begin{equation}
\underset{\overline{q} \in \Rset_+^n }{\text{minimize}} \quad \sum_{i=1}^n \widetilde{G}_i (\overline{q}_i) \quad \ \text{subject to} \quad \overline{q} \in \Xcal_0, \label{opt:equilibrium_q}
\end{equation}
where  the \emph{modified nodal cost functions} $\{\widetilde{G}_i (\overline{q}_i)\}_{i=1}^n$ are defined as
\begin{align}
\widetilde{G}_i (\overline{q}_i) := \begin{cases}
\displaystyle \int_{0}^{\overline{q}_i} g_i (z) \mathrm{d} z,  & \text{if } N_i > 0 \\
0,  & \text{if } N_i = 0.
\end{cases} \label{eq:modified_cost_q}
\end{align}
 The function $g_i (z)$ is defined according to  Eq. \eqref{eq:def_gi(z)}.
\end{enumerate}
\label{prop:equilibrium}
\end{propositio}

Several remarks are in order.  First, Proposition \ref{prop:equilibrium} implies that, although there may exist a multiplicity of Nash equilibria, the production profile that results at a Nash equilibrium is \emph{unique}.
Furthermore, Proposition \ref{prop:equilibrium}  provides an approach to the tractable calculation of the unique production profile at a Nash equilibrium via the solution of two finite-dimensional convex programs.
That is, one first solves problem \eqref{opt:equilibrium_q} for the unique nodal supply profile $q$ at a Nash equilibrium; and then solves problem \eqref{opt:equilibrium} to determine the unique production quantity of each producer. 
It is also worth noting that the  modified nodal cost functions $\{\widetilde{G}_i(\cdot)\}_{i=1}^n$  admit  closed-form expressions for a large family  of production cost functions $C_j(\cdot)$, e.g.,  piecewise-quadratic  functions.

We also note that Proposition \ref{prop:equilibrium} builds upon and generalizes existing results from the literature, \cite[Thm. 1]{Johari2011} and \cite[Thm 4.1]{Xu2014}, to accommodate the more general setting in which there are  transmission capacity constraints between producers and consumers. We conclude this subsection with a brief  discussion of the key ideas used in proving Proposition \ref{prop:equilibrium}.
The crux of  the derivation relies on the suitable design of cost functions to ensure equivalence between the stationarity conditions for problems  \eqref{opt:equilibrium} and \eqref{opt:equilibrium_q}  and  the best response conditions of all producers and the ISO at Nash equilibrium. This enables the characterization of the set of Nash equilibria  according to the optimal solution sets of the convex programs  \eqref{opt:equilibrium} and \eqref{opt:equilibrium_q}---a technique that is closely related to the use of potential functions in characterizing Nash equilibria in potential games \cite{monderer1996potential}.

More specifically, given a nodal supply profile $q$, the modified cost functions $\{\widetilde{C}_j (\cdot)\}_{j=1}^N$ are constructed in such a manner as to reflect the best response conditions for all producers according to the stationarity conditions (Eq. \eqref{eq:NEprice_1}--\eqref{eq:NEprice_2}) associated with problem \eqref{opt:equilibrium}.
Since the modified cost functions are parametric in the ISO's  decision $q$, we construct another optimization problem \eqref{opt:equilibrium_q} to enable the computation of the nodal supply profile $q$ at a Nash equilibrium. Towards this end, the modified nodal cost functions $\{\widetilde{G}_i (\cdot )\}_{i=1}^n$ are designed to ensure the equivalence between the stationarity conditions for problem \eqref{opt:equilibrium_q} and a certain fixed point condition, which expresses the ISO's decision $q$  as a best response to the best response of producers given the nodal supply profile $q$.
We refer the reader to  Appendix \ref{app:pf:prop:equilibrium} for the complete proof of Proposition \ref{prop:equilibrium}.

\subsection{Bounding the Efficiency Loss} \label{sec:effloss}

In Theorem \ref{thm:PoA_bound}, we provide an upper bound on the worst-case allocative efficiency loss  incurred at a Nash equilibrium. The bound sheds light on the explicit role of \emph{market structure} in determining the impact of producers' strategic behavior on market (in)efficiency at equilibrium.

\begin{theore}[Price of Anarchy] \label{thm:PoA_bound}
Let Assumptions \ref{ass:cost}-\ref{ass:cap} hold.  The price of anarchy (PoA) associated with the game $(\Ncal, \Xcal, \pi)$  satisfies
\begin{align} \label{eq:PoAmainbound}
\poa  \leq   1 + \max_{j \in \{ 1, \dots, N \}} \left\{ \frac{\ms_j}{\rsi_j - 1} \right\}.
\end{align}
\end{theore}

Several important consequences  can be deduced from Theorem \ref{thm:PoA_bound}---the most of important of which relate to the explicit role played by market structure in determining  the efficiency loss incurred at a Nash equilibrium. For instance, the PoA bound in \eqref{eq:PoAmainbound}  reveals the inherent limitation of the \emph{market share index}, by itself,  as an accurate predictor of market power;  and, provides a theoretical basis for the empirically observed effectiveness of  the \emph{residual supply index} in predicting the actual exercise of market power (as measured by price markups relative to perfectly competitive levels \cite{sheffrin2002predicting, swinand2010modeling, newbery2005market}). In particular, the PoA bound ensures a low efficiency loss for electricity markets  in which all participating power producers have \emph{large} residual supply indices---irrespective of their market shares. 
It is also worth mentioning that, when all producers are concentrated at a single node, we recover as a special case the PoA bound established by Xu et al. \cite{Xu2014} for single-node electricity markets.

Additionally, we note that the PoA bound in \eqref{eq:PoAmainbound} hints at the possibility of a Braess-like paradox, where an increase in a network's transmission capacity can result in the  (counterintuitive) increase in the aggregate cost of generation at Nash equilibrium. The argument behind such a claim is that an increase in a  transmission line's capacity can lead to an increase in the maximum nodal supply  at certain nodes in the network. This, in turn, may increase the efficiency loss at Nash equilibrium, as the PoA bound is non-decreasing in the maximum nodal supply of each node. 
In Section \ref{sec:numerical}, we examine a two-node network, and  establish a necessary and sufficient condition (cf. Lemma \ref{lem:NE_2node}) under which the strengthening of the network's transmission line capacity  results in this seemingly paradoxical behavior. We conclude this subsection with a proof of Theorem \ref{thm:PoA_bound}.

\begin{proof}[Proof of Theorem \ref{thm:PoA_bound}]
Let $(q, \theta)$ be a Nash equilibrium. Using the assumption that each producer's  cost function $C_j (x_j)$ is strictly positive and strictly increasing over $(0, \infty)$, we have that
\begin{align}
C_j (x_j) &\leq C_j (x_j) + \frac{x_j C_j (x_j) - \int_0^{x_j} C_j (z) \mathrm{d}z}{\sum_{k \in \Ncal_i, k \neq j} X_k - q_i } \label{eqn:pf_PoA_0}\\
& = \widetilde{C}_j \left( x_j; q_i \right) \\
&\leq C_j (x_j) \left( 1 + \frac{x_j}{\sum_{k \in \Ncal_i, k \neq j} X_k - q_i } \right). \label{eqn:pf_PoA_1}
\end{align}
Let $x^*$ be an efficient production profile. It holds that:
\begin{align}
 \nonumber &\sum_{i = 1}^n \sum_{j \in \Ncal_i} C_j (x_j (q, \theta) ) \\
 & \qquad \leq \sum_{i = 1}^n \sum_{j \in \Ncal_i} \widetilde{C}_j (x_j (q, \theta ); q_i ) \label{eqn:pf_PoA_2}\\
& \qquad \leq    \sum_{i = 1}^n \sum_{j \in \Ncal_i} \widetilde{C}_j (x_j^*; q_i ) \label{eqn:pf_PoA_3} \\
& \qquad \leq \sum_{i = 1}^n \sum_{j \in \Ncal_i} C_j (x_j^*)  \left( 1 + \frac{x_j^*}{\sum_{k \in \Ncal_i, k \neq j} X_k - q_i } \right) \label{eqn:pf_PoA_4} \\
\begin{split}
&\qquad \leq \sum_{i = 1}^n \sum_{j \in \Ncal_i} C_j (x_j^*) \cdot \\
 & \qquad \left( 1 + \max_{j \in \Ncal_i, i \in \Vcal} \left\{ \frac{\min \left\{ X_j, q_i^{\max} \right\}}{ \left( \sum_{k \in \Ncal_i, k \neq j} X_k \right) - q_i^{\max} } \right\} \right).
\end{split} \label{eqn:pf_PoA_5}
\end{align}
Here, inequality \eqref{eqn:pf_PoA_2} follows from \eqref{eqn:pf_PoA_0}; inequality \eqref{eqn:pf_PoA_3} follows from the optimality of $x (q, \theta)$ for problem \eqref{opt:equilibrium}; inequality \eqref{eqn:pf_PoA_4} follows from \eqref{eqn:pf_PoA_1}; and, inequality  \eqref{eqn:pf_PoA_5} follows from the feasibility of $x^*$ and $q$.  It follows from inequality \eqref{eqn:pf_PoA_5} that the PoA at a Nash equilibrium is upper bounded by
\begin{align*}
\poa  \leq 1 + \max_{j \in \Ncal_i, i \in \Vcal} \left\{ \frac{\min \left\{ X_j, q_i^{\max} \right\}}{ \left( \sum_{k \in \Ncal_i, k \neq j} X_k \right) - q_i^{\max} } \right\}.
\end{align*}
The desired result follows,  as  $\ms_j = \min \left\{ X_j, q_i^{\max} \right\} /  q_i^{\max} $ and $\rsi_ j = ( \sum_{k \in \Ncal_i, k \neq j} X_k )/ q_i^{\max} $  by definition.
\end{proof}

\subsection{Bounding the Price Markups} \label{sec:markup}
In this section, we derive  upper bounds on the nodal price markups at a Nash equilibrium in terms of the \emph{Lerner index}---a standard ex post measure of market power \cite{lerner1934concept}.

\begin{definitio}[Lerner Index]
Given a strategy profile $(q, \theta) \in \Xcal$, the \emph{Lerner index} of producer $j \in \Ncal_i$ at  node $i \in \Vcal$
is defined as
\begin{align*}
\lerner_j (q, \theta) := \frac{p_i (q, \theta) - \partial^+ C_j (x_j (q, \theta) ) / \partial x_j  }{p_i (q, \theta)}.
\end{align*}  
\end{definitio}
It is straightforward to show that, under the additional assumption that $(q, \theta) \in \Xcal_{\nash}$, the Lerner index of each producer $j$ is guaranteed to satisfy $\lerner_j (q, \theta) \in [0, 1]$. Accordingly, a  Lerner index close to one (resp. zero) implies a large (resp. small) price markup relative to the producer's true marginal cost.
The following corollary to Proposition \ref{prop:equilibrium} shows that the  Lerner index of each producer is upper bounded at equilibrium.

\begin{corollar} \label{cor:Lerner}
Let Assumptions \ref{ass:cost}-\ref{ass:cap} hold. At   a Nash equilibrium $(q, \theta)$,
 the Lerner index of  producer $j \in \Ncal_i$  satisfies 
\begin{align}
\lerner_j (q, \theta) \leq  \frac{\ms_j}{\ms_j + \rsi_j -1} \label{eq:LImainbound}
\end{align}
if $x_j (q, \theta) < X_j$. Furthermore, if producer $j \in \Ncal_i$ has a   differentiable cost function and $x_j (q, \theta) = q_i^{\max} < X_j$, then 
\begin{align}
\lerner_j (q, \theta) = \frac{1}{\rsi_j}. \label{eq:LI_tightness}
\end{align}
\end{corollar}

Corollary \ref{cor:Lerner} reveals the potential for large price markups at nodes that have a dominant producer with a small residual supply index. We provide empirical evidence in support of this  claim  in Section \ref{sec:empirical} using historical spot price data from the Great Britain Electricity Pool.  

We also note that, in a related line of investigation, Newbery \cite{newbery2009predicting} employs a Cournot oligopoly model to characterize a producer's Lerner index in terms of its RSI under a variety of assumptions on the market structure. 
In particular, under the assumption of symmetric producers, Newbery shows the Lerner index of each producer  to be inversely proportional to its RSI.
This functional relationship is similar in structure  to our  upper bound on the Lerner index in Corollary \ref{cor:Lerner}, despite the differences in the underlying models of competition employed.
We note, however, that implicit in the Cournot model, which Newbery treats, is the requirement of nonzero demand  elasticity. Consequently, his results cannot be applied to the  setting considered in this paper, which considers a perfectly inelastic demand model.

\begin{proof}[Proof of Corollary \ref{cor:Lerner}]  Let $(q, \theta)$ be a Nash equilibrium. It follows from condition \eqref{eq:NEprice_1} in Proposition \ref{prop:equilibrium} that 
\begin{align*}
p_i (q, \theta)   \leq \frac{\partial^+ \widetilde{C}_j (x_j (q, \theta); q_i)}{\partial x_j}  
\end{align*}
if $x_j (q, \theta) < X_j$.  Calculating the right derivative of the modified cost function yields
\begin{align} \label{eq:mark1}
 p_i (q, \theta) \leq \left( 1 + \dfrac{x_j (q, \theta)}{\sum_{k \in \Ncal_i \setminus \{j\}}  X_k  - q_i} \right)\frac{  \partial^+  C_j (x_j (q, \theta))}{\partial x_j} .
\end{align}
It follows that the Lerner index of producer $j$ satisfies
\begin{align*}
\lerner_j (q, \theta) &= 1 - \frac{\partial^+ C_j (x_j (q, \theta) ) / \partial x_j  }{p_i (q, \theta)} \\
& \leq 1 - \frac{1}{ 1 +  x_j (q, \theta)\left/ \left(  \sum_{k \in \Ncal_i \setminus \{j\}}  X_k  - q_i \right) \right.} \\
& = \frac{x_j (q, \theta) }{ x_j (q, \theta) +   \sum_{k \in \Ncal_i \setminus \{j\}}  X_k  - q_i } .
\end{align*}
The Lerner index bound in \eqref{eq:LImainbound} follows, as it necessarily holds that $q_i \leq q_i^{\text{max}}$ and $x_j (q, \theta) \leq \min\{X_j, q_i^{\text{max}}\}$. 

Additionally, it follows from condition \eqref{eq:NEprice_1} in Proposition \ref{prop:equilibrium} that inequality \eqref{eq:mark1} holds with equality if producer $j$'s cost function $C_j$ is differentiable at $x_j (q, \theta)$. This implies the satisfaction of Eq. \eqref{eq:LI_tightness} if $x_j (q, \theta) = q_i^{\text{max}} < X_j$.
\end{proof}

\begin{figure*}
\centering
\includegraphics[width = 0.75 \linewidth]{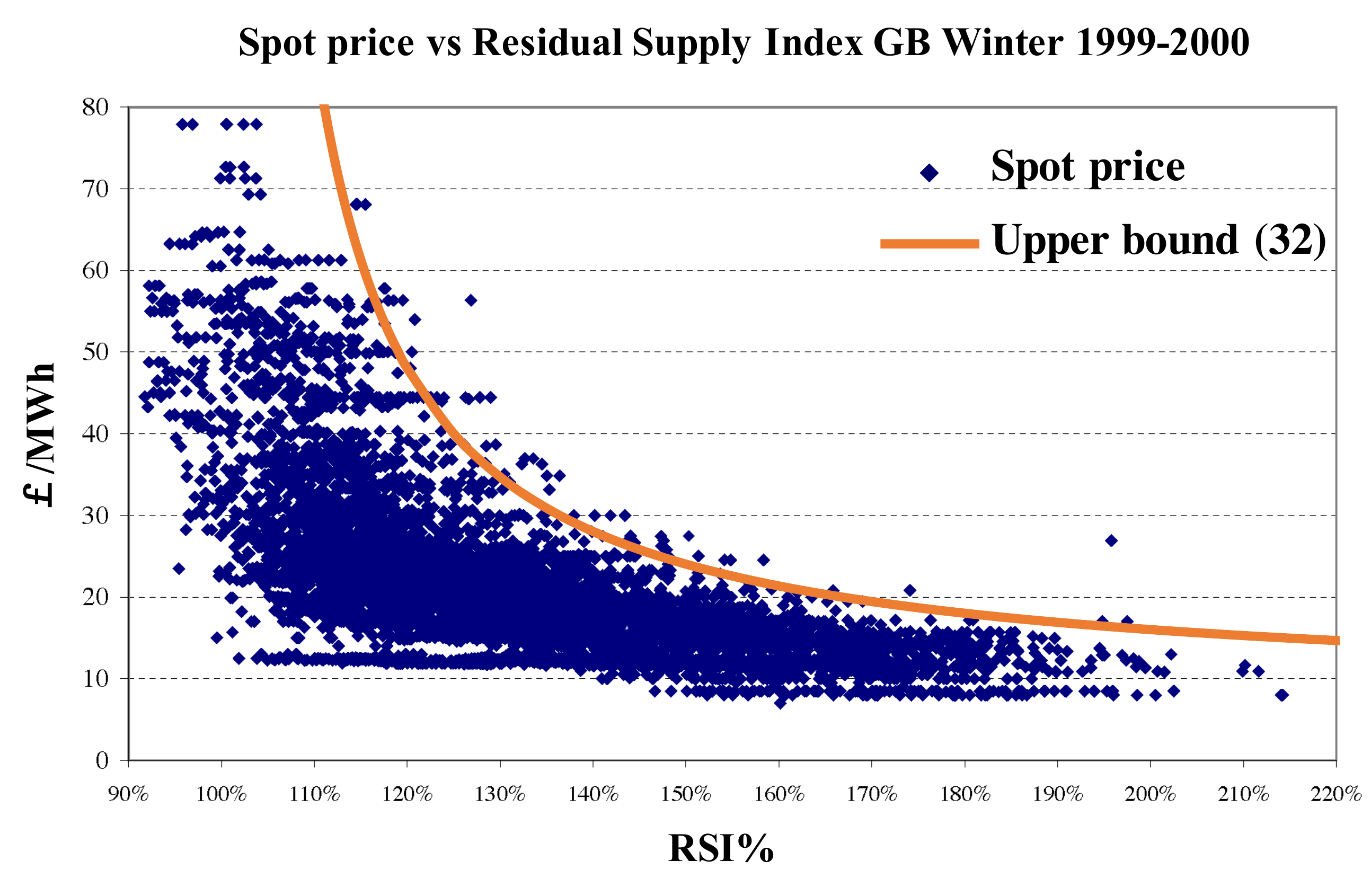}
\caption{A scatter plot obtained from \cite{newbery2005market} of spot electricity prices versus the residual supply index (RSI) of the  generator with the largest uncontracted production capacity in the Great Britain Electricity Pool during the 1999-2000 winter. The solid line corresponds to the price upper bound  \eqref{eq:estimated}.}
\label{fig:newbery}
\end{figure*}

\subsection{Comparison to Historical Market Data} \label{sec:empirical}

It follows from \eqref{eq:mark1} that---at a Nash equilibrium $(q, \theta)$---the price at any node $i \in \Vcal$ is upper bounded by 
\begin{align} \label{eq:markup}
p_i (q, \theta) \leq \left( 1 + \dfrac{\ms_j}{\rsi_j -1} \right)\frac{  \partial^+  C_j (x_j (q, \theta))}{\partial x_j} ,
\end{align}
given any  producer $j \in \Ncal_i$ whose production capacity constraint is nonbinding at Nash equilibrium.\footnote{We note that one such producer is guaranteed to exist at each node given the satisfaction of Assumption \ref{ass:cap}, i.e., that there is \emph{no pivotal producer} in the market.} In this section, we evaluate the predictive accuracy of the  price bound \eqref{eq:markup} using market data  from the Great Britain (GB) Electricity Pool for the winter of 1999-2000. During this time period, producers in the GB Electricity Pool were required to bid supply functions
in the form of nondecreasing step functions, and were remunerated for their cleared production quantities according to a uniform market clearing price set at the pool level. See \cite{simmonds2002regulation, offer1998} for a detailed description of the underlying market mechanism in place in Great  Britain during this time period. 
Figure \ref{fig:newbery} contains a scatter plot obtained from  \cite{newbery2005market} of  spot electricity prices (\pounds /MWh) versus the residual supply index  of the producer with the largest uncontracted production capacity during this time period.

In order to calculate the price bound \eqref{eq:markup}, we require estimates of the underlying producer's  \emph{true marginal cost} and \emph{market share}. If we adopt a simplifying assumption of \emph{linearity} of the producer's cost function, then the spot prices in  Figure \ref{fig:newbery} imply an upper bound  on the producer's true marginal cost of \pounds 8/MWh.  
As for the producer's market share, we do not have access to information on the  producer's uncontracted production capacity  or  detailed  demand data for the time period under consideration.  We, therefore, estimate the producer's market share according to  a worst-case upper bound of  $\ms \leq 1$.  The combination of these two approximations implies an upper bound on the spot prices of the form
\begin{align} \label{eq:estimated}
\text{spot price} \leq \left(  \dfrac{\rsi}{\rsi -1} \right) 8 \quad \text{(\pounds/MWh)},
\end{align}
which is valid for all $\rsi \in (1, \infty)$. 

We plot the estimated upper bound \eqref{eq:estimated} against the historical market data in Figure \ref{fig:newbery}. Notice that, with the exception of a few  outliers in the data, the a priori upper bound \eqref{eq:estimated} and the  upper envelope of  the observed spot prices are in near agreement.  The near agreement between our theoretical bound and the spot price data is particularly striking in light of the apparent discrepancy between the class of piecewise-constant supply functions employed in the GB Electricity Pool  and the  scalar-parameterized family considered in this paper. 
This empirical observation lends some credence to the claim that the parametric family of supply functions studied in this paper---although stylized in nature---preserves the key structural determinants of market power and the mechanisms by which market power is exercised, e.g., through the economic withholding of capacity when (residual) supply is scarce.
\section{A Transmission Expansion Paradox} \label{sec:case_study}

In this section, we restrict our attention to the  setting of a two-node power network, and provide a characterization of market structures under which a \emph{Braess-like  paradox} emerges due to the exercise of market power. Specifically, we characterize a range of  scenarios in which the strengthening of the network's transmission line capacity  results in the counterintuitive \emph{increase} in the total cost of generation at the Nash equilibrium.\footnote{Braess's original paradox revealed  that the addition of new roads to a traffic network can manifest in the counterproductive effect of increasing drivers' total commute time at equilibrium \cite{braess1968paradoxon}.}

\subsection{A Two-node Network} \label{sec:case_study_system}

Consider a two-node power network with a nodal demand profile given by $d = \left(d_1, d_2 \right) > 0$, and a total demand of $D = d_1 + d_2$. We denote the capacity of the transmission line connecting the two nodes by  $c \in \Rset_+$. 
We assume that the producers that are common to a node are \emph{symmetric} and have \emph{linear cost} functions.
Specifically, we denote the production capacity and cost function of each  producer $j \in \Ncal_i$ at node $i \in  \{1, 2 \}$ by $X_j = \capx_i$ and $C_j (x_j) = (\beta_i x_j)^+$, respectively.
We enforce the satisfaction of Assumption \ref{ass:cap} (i.e., that  no producer be pivotal) by requiring that
\begin{align*}
 \frac{\capx_i (N_i - 1)}{D} > 1
\end{align*}
for each node $ i \in \{1,2\}$. 
This condition, in combination with the assumption of linear production costs, guarantees the existence of a \emph{unique} Nash equilibrium for the networked supply function game.\footnote{More specifically, the linearity of production costs implies the differentiability of each modified cost function $\widetilde{C}_j$. This, in combination with statement (iii) in Proposition \ref{prop:equilibrium},  guarantees the uniqueness of the LMPs at a Nash equilibrium. The uniqueness of both the LMPs and the production profile at Nash equilibrium implies the uniqueness of Nash equilibrium.}
Finally, we assume that $ \beta_2 > \beta_1$, i.e.,  that node 2 is more expensive than node 1.
It follows that the efficient production cost can be calculated as
$$\cost_{\text{Eff}} := \sum_{j=1}^N C_j(x_j^*) =  \beta_1 D + (\beta_2 - \beta_1) (d_2 - c)^+.$$

\subsection{Economic Inefficiency of Transmission Expansion} \label{sec:braess}
In what follows, we derive a necessary and sufficient condition on the market structure under which transmission capacity expansion results in the increase in the total cost of generation at Nash equilibrium.
We first characterize  the locational marginal prices (LMPs) and nodal supply profile that result at the Nash equilibrium for the system under consideration. Let $q = (q_1, q_2)$ be the nodal supply profile at Nash equilibrium. It follows from Proposition \ref{prop:equilibrium} that $q$ is  the unique optimal solution to the following convex program:

\begin{equation}
\begin{alignedat}{8}
& \underset{\overline{q} \in \Rset_+^2}{\text{minimize}} \quad && \sum_{i=1}^2 \int_0^{\overline{q}_i} \beta_i \left( 1 +  \frac{1 / N_i }{   (N_i - 1) \capx_i / z- 1  } \right) \mathrm{d} z \\
& \text{subject to} \quad && \overline{q}_1 + \overline{q}_2 = D \\
&&& |\overline{q}_1 - d_1 | \leq c.
\end{alignedat}\label{opt:2node}
\end{equation}
Additionally, the LMPs at the unique Nash equilibrium can be computed according to Eq. \eqref{eq:stationarity_q3} in Appendix \ref{app:pf:prop:equilibrium}. More specifically, if there is positive production at node 2 at the Nash equilibrium, i.e., $q_2 > 0$, then the corresponding LMP at each node $i \in \{1, 2\}$ is given by
\begin{align}
p_i = \beta_i\left( 1+ \left(N_i  \left(   \dfrac{K_i (N_i - 1)}{q_i } - 1 \right) \right)^{-1} \right) . \label{eq:p_2node_case1}
\end{align}
 If instead $q_2 = 0$, then the corresponding LMPs at each node are identical and are given by
\begin{align}
p_1 = p_2 = \beta_1\left( 1  +  \left(N_1 \left(   \dfrac{K_1(N_1- 1)}{D} - 1 \right) \right)^{-1} \right). \label{eq:p_2node_case2}
\end{align}

We let $\cost_{\nash}$ denote the aggregate production cost at Nash equilibrium. It is given by
\begin{align*}
\cost_{\nash} = \beta_1 q_1 + \beta_2 q_2.
\end{align*}
In what follows, we  investigate the behavior of the aggregate production cost at Nash equilibrium as a function of the  network's transmission capacity $c$. 
In particular, we provide an explicit characterization of the right derivative of $\cost_{\nash}$ with respect to $c$, and establish a necessary and sufficient condition under which this derivative is guaranteed to be strictly positive---that is, a condition under which the aggregate production cost at the Nash equilibrium \emph{increases}  with the transmission capacity.

\begin{lemm} \label{lem:NE_2node} Let $(p_1,p_2)$ denote the LMPs at the Nash equilibrium. Then
\begin{align}
\frac{\partial^+ \cost_{\nash}}{\partial c} =  (\beta_2 - \beta_1) \cdot \sign (p_1 - p_2),\label{eq:cost_NE_derivative}
\end{align}
where $\sign (0) = 0$. Additionally,  $p_1 > p_2$ 
if and only if 
\begin{equation}
\dfrac{ 1 +   \left(N_1 \left(   \dfrac{K_1(N_1 - 1)}{d_1- c } - 1 \right)\right)^{-1}}{ 1 +  \left( N_2  \left(   \dfrac{K_2 (N_2 - 1)}{d_2 +c } - 1 \right)\right)^{-1} } > \dfrac{\beta_2}{\beta_1}.
\label{eq:braess}
\end{equation}
\end{lemm}
The proof of Lemma \ref{lem:NE_2node} is deferred to Appendix \ref{app:pf:lem:NE_2node}.
The  necessary and sufficient condition  in \eqref{eq:braess} sheds light on the role of market and network structures in driving the emergence of this Braess-like paradox.  Loosely speaking,  if the network's transmission capacity is sufficiently limited, and if the market power of producers (as measured by the residual supply index) at node 1 is  sufficiently large relative to the market power of producers at node 2, then the  Nash equilibrium will result in LMPs that correspond to the (inefficient) transmission of power from the \emph{high-marginal cost node} 2 to the \emph{low marginal cost node} 1.  For such market structures, a small increase in the network's transmission capacity will induce an increase (resp. decrease) in the production at node 2 (resp. node 1), thereby increasing the aggregate production cost at Nash equilibrium.

It is  worth noting that Sauma and Oren \cite{sauma2006proactive}  uncover an example of a similar transmission expansion paradox in the context of a simultaneous-move networked Cournot model. In contrast to their characterization, which is entirely numerical in nature, Lemma \ref{lem:NE_2node} sheds light on how  market structure might induce market power that  gives rise to such counterintuitive market outcomes under transmission expansion.

It is also possible to extend this line of reasoning to establish similar conditions under which an increase in `competition' at a node 2 (as measured by the number of producers at that node) results in the increased dispatch of the
high marginal cost generation at node 2, in place of the lower marginal cost generation at node 1---thereby increasing the total cost of generation at the Nash equilibrium. This seemingly paradoxical behavior is in direct contrast to the more commonly held belief that the market entry of additional producers serves to improve economic efficiency, in general.
We note that Berry et al. first described such counterintuitive `network effects' in their seminal paper \cite{berry1999understanding}, which employs  a computational approach to the  calculation of linear supply function equilibria in constrained transmission systems.

\subsection{Numerical Analysis} \label{sec:numerical}

We consider a two-node power network with a nodal demand profile given by $d_1 = d_2 = 1$, and set the  number of producers at nodes 1 and 2 to be $N_1 = 3$ and $N_2 = 10$, respectively. 
We  fix the marginal cost of producers at node 1 to be $\beta_1 = 1$, and vary the marginal cost of producers at node 2 between two values $\beta_2 \in  \{1.15, 1.45\}$. All producers are assumed to have identical production capacities $\capx_1 = \capx_2 = 0.51D$.


\begin{figure}[http]
\centering

\begin{subfigure}{0.48\linewidth}
\centering
$\beta_2 = 1.15$ \vspace{.04in}

\includegraphics[width = \linewidth]{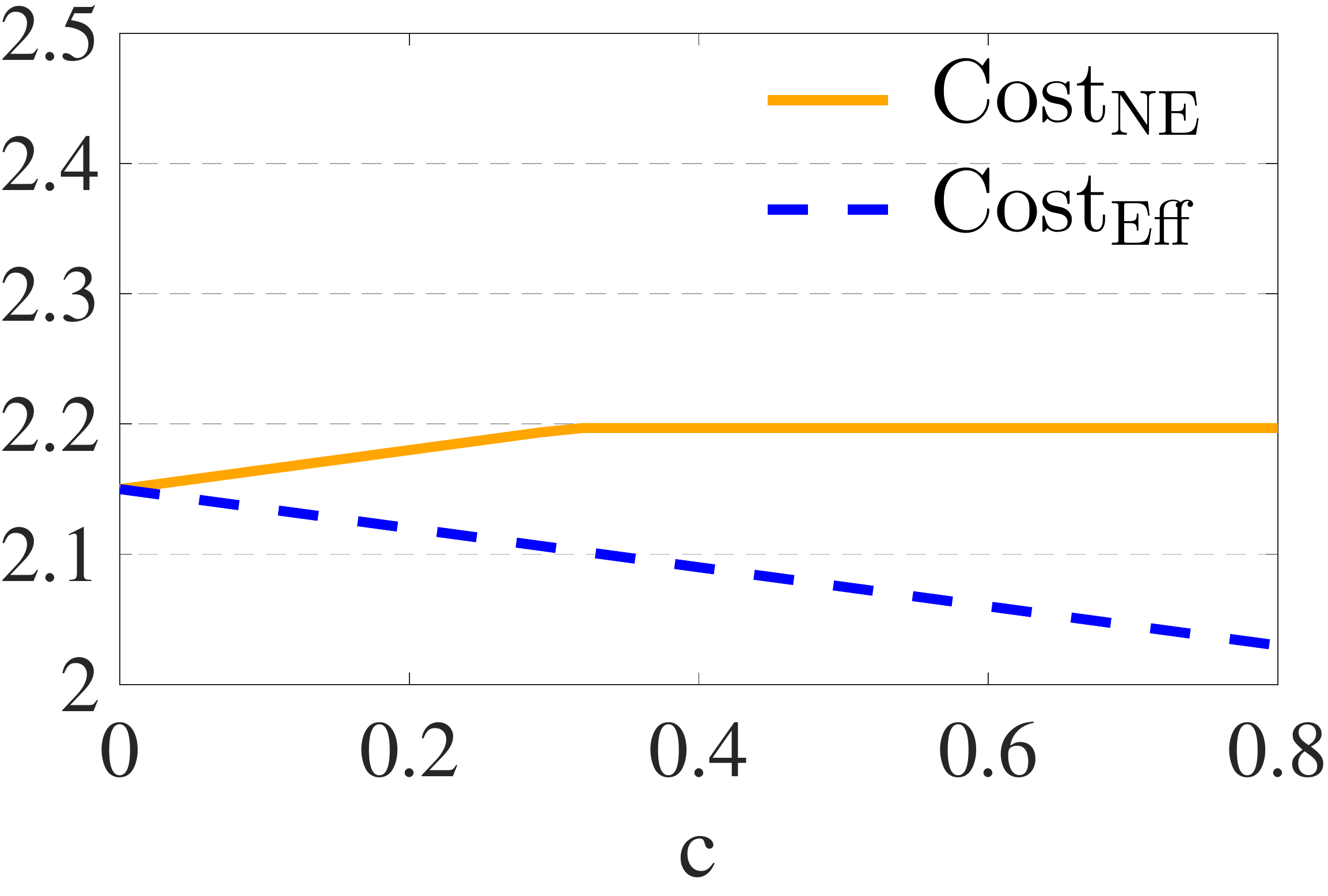}
\caption{}
\label{fig:cost_c1}
\end{subfigure}
\begin{subfigure}{0.48\linewidth}
\centering
$\beta_2 = 1.45$ \vspace{.04in}

\includegraphics[width = \linewidth]{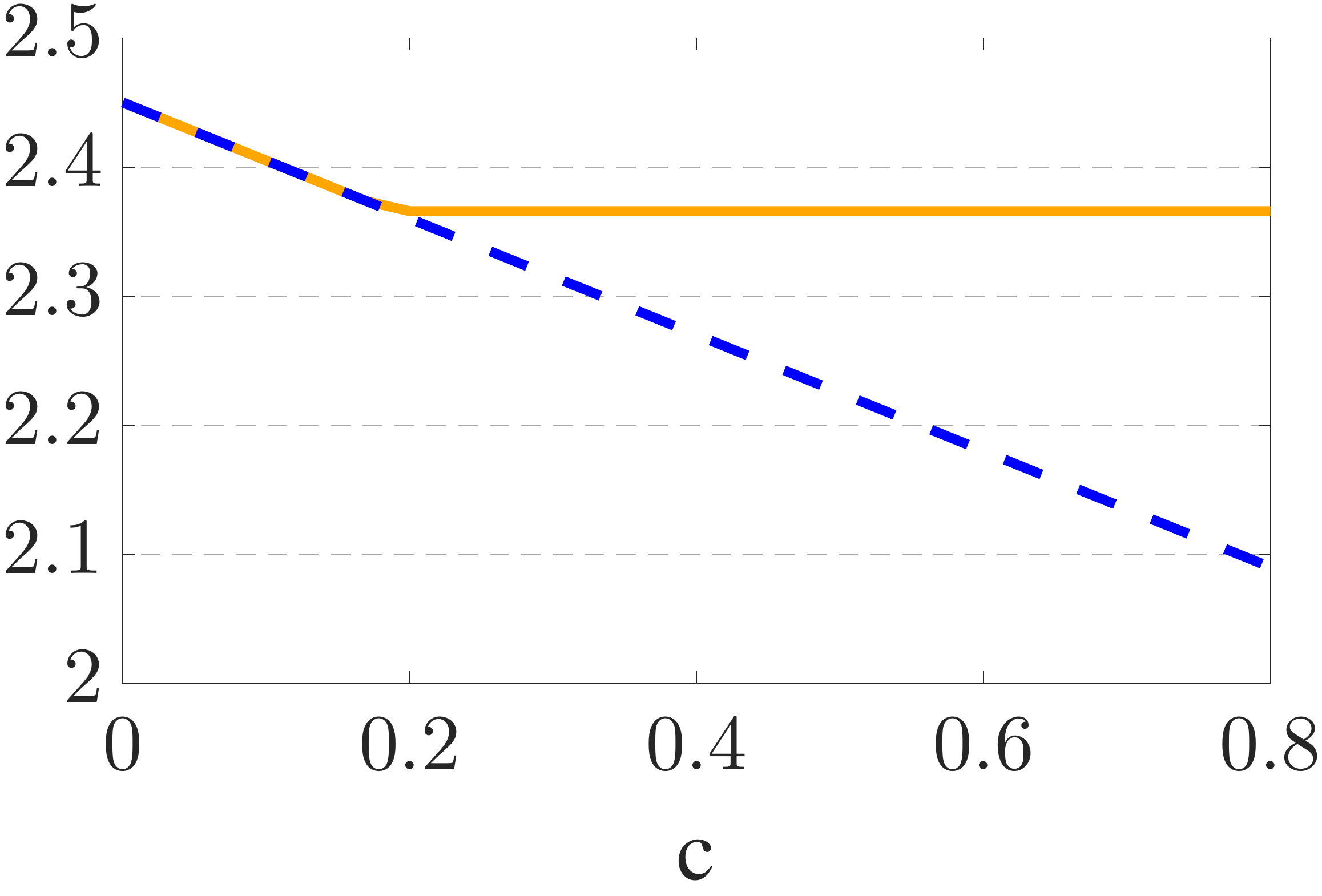}
\caption{}
\label{fig:cost_c2}
\end{subfigure}

\vspace{.1in}

\begin{subfigure}{0.48\linewidth}
\centering
\includegraphics[width = \linewidth]{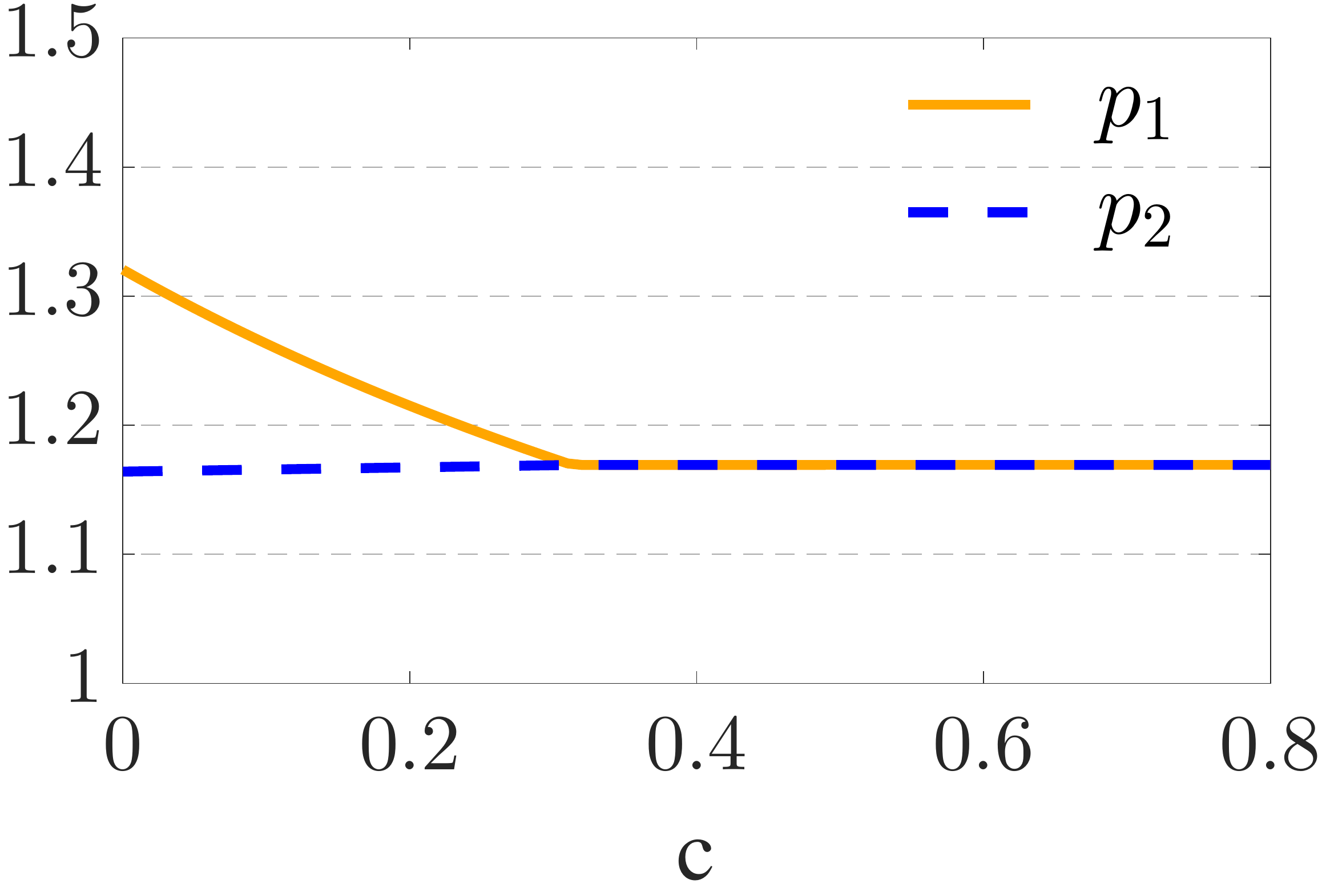}
\caption{}
\label{fig:p_c1}
\end{subfigure}
\begin{subfigure}{0.48\linewidth}
\centering
\includegraphics[width = \linewidth]{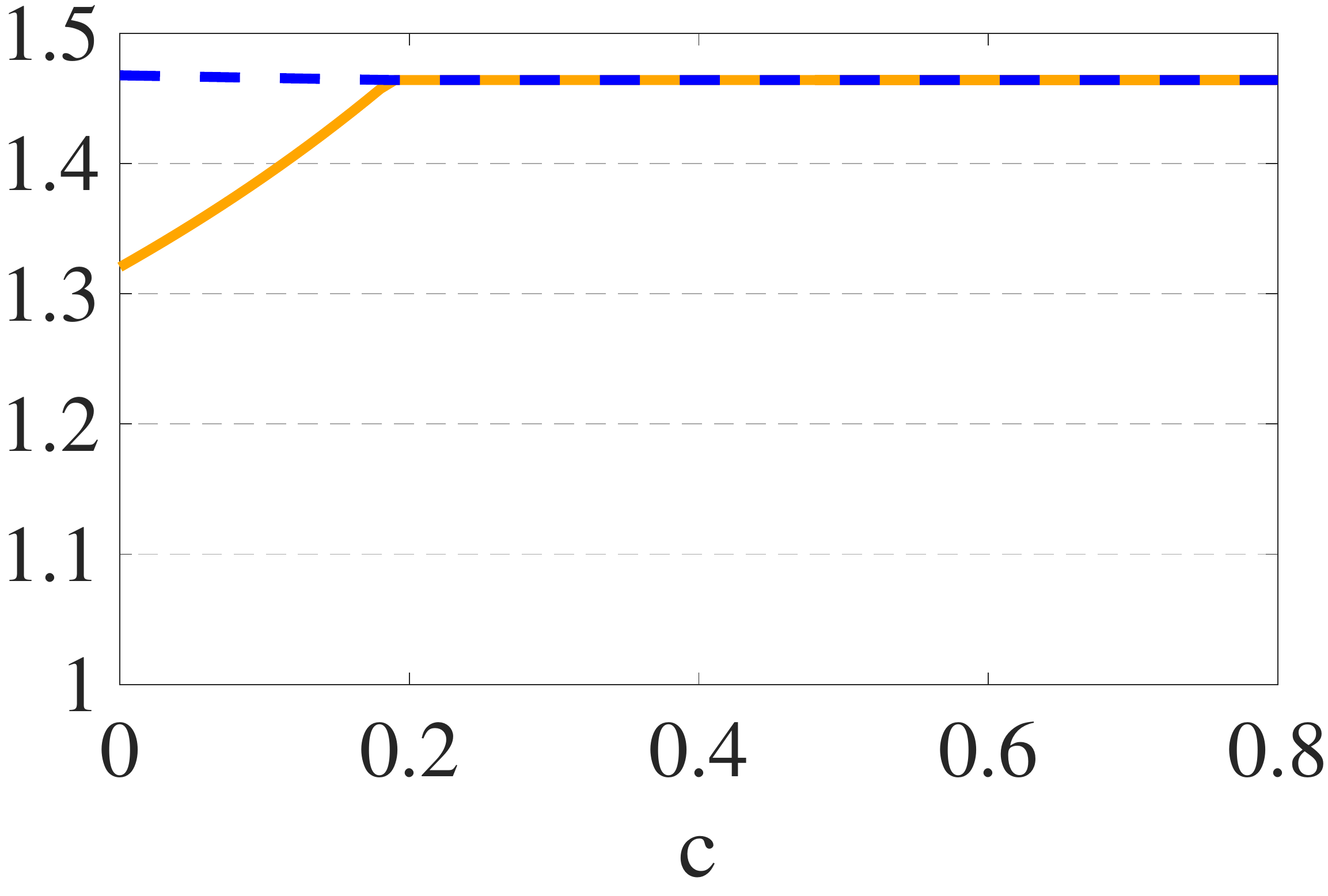}
\caption{}
\label{fig:p_c2}
\end{subfigure}

\caption{We fix two different values of $\beta_2$, and vary the transmission capacity $c$ from 0 to 0.8.  
Figures \ref{fig:para}(\subref{fig:cost_c1})-\ref{fig:para}(\subref{fig:cost_c2}) plot the efficient cost and the aggregate production cost at Nash equilibrium. Figures \ref{fig:para}(\subref{fig:p_c1})-\ref{fig:para}(\subref{fig:p_c2}) plot the nodal prices at Nash equilibrium.}
\label{fig:para}
\end{figure}


The leftmost plots in Figure \ref{fig:para} correspond to a marginal cost of $\beta_2 = 1.15$ at node 2. For this choice of marginal cost, it is straightforward to show that condition \eqref{eq:braess} is satisfied for all  transmission capacity values $c \in [0, 0.3]$. Indeed, Figure \ref{fig:para}(\subref{fig:cost_c1}) shows the aggregate production cost at Nash equilibrium to be strictly increasing over this range of capacity values, and constant for all other capacity values greater than 0.3.   
It is also worth noting that  Figure \ref{fig:para}(\subref{fig:p_c1}) corroborates the necessary and sufficient condition  \eqref{eq:braess}, as the nodal prices  satisfy  $p_1 > p_2$ for all transmission capacities less than or equal to 0.3
The rightmost plots in Figure \ref{fig:para} correspond to a marginal cost of $\beta_2 = 1.45$ at node 2. It is straightforward to show that, for this choice of marginal cost,  condition \eqref{eq:braess} is violated for all transmission capacities $c \geq 0$. Accordingly, Figure \ref{fig:para}(\subref{fig:cost_c2}) reveals the aggregate production cost to be monotone nonincreasing in the network's transmission capacity.

\section{Conclusion} \label{sec:conclusion}

We conclude the paper with a brief discussion surrounding possible  directions for future research.
\emph{First}, the equilibrium analysis of the supply function game considered in this paper relies on the assumption that the nodal demand profile is both inelastic and known.
It would be of interest to generalize our analysis to the setting in which the demand  exhibits price elasticicity and/or randomness in the values it takes; and quantify the extent to  which uncertainty and price elasticity of demand  serves in mitigating the exercise of market power by strategic power producers---in a similar spirit to prior analyses of supply function equilibria in the absence of network constraints \cite{Klemperer1989, Baldick2004}.
\emph{Second}, our analysis relies on the simplifying assumption that producers and the ISO choose their strategies simultaneously. 
Such an assumption facilitates the tractability of equilibrium analysis, which, in turn,  provides structural insights on the influence of generator capacity and  transmission constraints on the ability of producers to exert market power. Nevertheless, it is well understood in the literature that the assumption of simultaneous movement between the ISO and producers---as compared to the more plausible sequential-move formulation---will manifest in market equilibria that \emph{underpredict}  the intensity  of competition in power networks with little to no transmission congestion\cite{Neuhoff2005, Yao2008}. 
It would, therefore, be  of interest to investigate  the design of solution concepts that better approximate the  sequential nature of the interaction between producers and the ISO, while preserving tractability of analysis.
As one possible starting point, it would be interesting  to analyze an alternative model of simultaneous movement in which the ISO's strategic variables are the nodal price differences, as opposed to nodal supply quantities. In the specific context of  networked Cournot models, such approximations have been previously shown to provide more accurate predictions of market outcomes in networks with little to no congestion---see, for example, Yao et al. \cite{Yao2008}.
\emph{Third}, our analysis of the supply function game is static in nature. As to whether or not these equilibria can be attained as the stable outcome of a natural learning dynamic remains unknown. \emph{Finally}, it would be of interest to test the predictive accuracy of the theoretical bounds established in this paper against more comprehensive market data drawn from LMP-based energy markets currently in operation.

\section*{Acknowledgment}

The authors would like to thank Subhonmesh Bose, Benjamin Hobbs, Shmuel Oren,  and Pravin Varaiya for helpful comments and discussion. 
This paper builds on the authors' preliminary results published as part of the 2016 IEEE Conference on Decision and Control (CDC) \cite{lin2016parameterized}.  
This work was supported in part by the National Science Foundation under  grants ECCS-1351621 and  IIP-1632124, the  
Department of Energy under the CERTS initiative, and the Simons Institute for the Theory
of Computing.

\bibliographystyle{IEEEtran}
\bibliography{reference}{\markboth{References}{References}}

\begin{IEEEbiography}[{\includegraphics[width=1in,height=1.25in,clip,keepaspectratio]{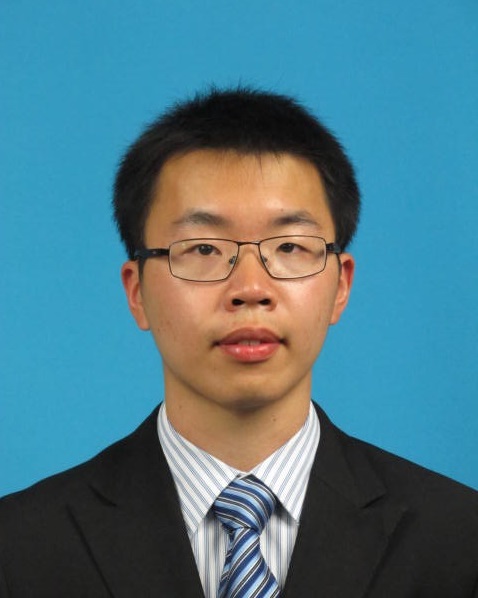}}]{Weixuan Lin} has been pursuing the Ph.D. degree in Electrical and Computer Engineering from Cornell University, Ithaca, NY, USA, since 2013. He received the B.S. degree in Electrical Engineering with high honors from Tsinghua University, Beijing, China, in 2013. 
His research interests include stochastic control, algorithmic game theory and online algorithms, with particular applications to the management of uncertainty in distributed energy resources and the design and analysis of electricity markets.
He is a recipient of the Jacobs Fellowship and the Hewlett Packard Fellowship. 
\end{IEEEbiography}

\begin{IEEEbiography}[{\includegraphics[width=1in,height=1.25in,clip,keepaspectratio]{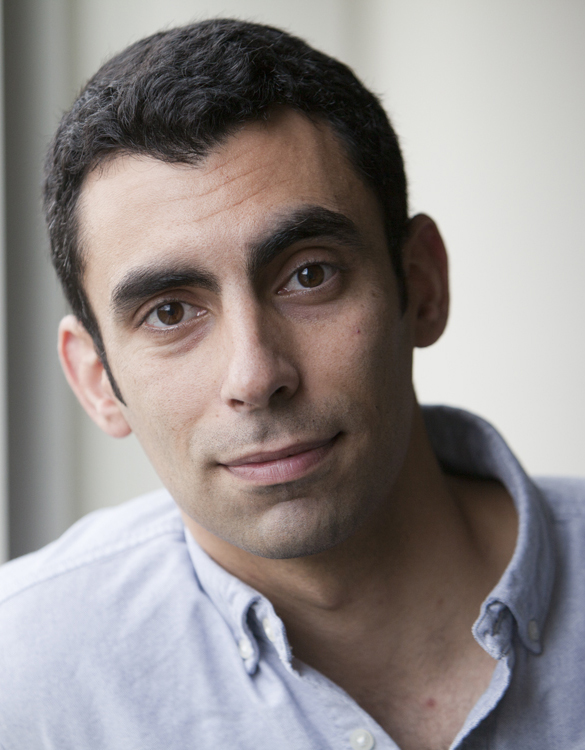}}]{Eilyan Bitar}  received the B.S. and Ph.D. degrees in Mechanical Engineering from the University of California at Berkeley in 2006 and 2011, respectively.

He currently serves as an Associate Professor and the David D. Croll Sesquicentennial Faculty Fellow in the School of Electrical and Computer Engineering at Cornell University in Ithaca, NY, USA.  Prior to joining Cornell in the Fall of 2012, he was engaged as a Postdoctoral Fellow in the department of Computing and Mathematical Sciences at the California Institute of Technology and at the University of California, Berkeley in Electrical Engineering and Computer Science, during the 2011-2012 academic year. 
His current research interests include stochastic control and game theory with applications to electric power systems and markets.

Dr. Bitar is a recipient of the NSF Faculty Early Career Development Award (CAREER), the John and Janet McMurtry Fellowship, the John G. Maurer Fellowship, and the Robert F. Steidel Jr. Fellowship.
\end{IEEEbiography}
\clearpage

\begin{appendices}


\section{Competitive Equilibrium} \label{sec:CE}

Under the assumption of price-taking behavior, we establish both the existence and the efficiency of \emph{competitive equilibria}, which we define as follows.

\begin{definitio}[Competitive Equilibrium] \label{def:CE}
The nodal price profile $p >0$ and producers' bid profile $\theta \geq 0$ constitute a \emph{competitive equilibrium} (CE) if  producers are \emph{maximizing their profits} given the nodal prices $p$,  i.e.,
\begin{align*}
\theta_j \in \argmax \left\{  \, p_i \cdot S_j \left( \overline{\theta}_j, p_i \right) - C_j \left( S_j \left( \overline{\theta}_j, p_i \right)  \right) \, \left| \, \overline{\theta}_j \geq 0  \right. \right\}
\end{align*}
for each   node $i \in \Vcal$ and  producer  $j \in \Ncal_i$; and  the \emph{market clears} at each node $i \in \Vcal$ according to
\begin{align*}
\sum_{j \in \Ncal_i} S_j (\theta_j, p_i)  = q_i,
\end{align*}
where the nodal supply profile $q$ solves the ISO's economic dispatch (ED) problem 
\begin{align*}
q \in \argmin \left\{ \left. \sum_{i=1}^n G_i (\overline{q}_i; \theta) \; \right| \; \overline{q} \in \Xcal_0 \right\}.
\end{align*}
\end{definitio}

At a competitive equilibrium, each producer maximizes its profit while taking its nodal price as given. In addition, the market clears at each node in the network according to a nodal price profile, which induces a nodal supply profile that solves the ISO's ED problem. 
Next, we show that competitive equilibria exist, and that any competitive equilibrium is efficient.

\begin{propositio}[Existence and Efficiency of CE] \label{thm:CE}
Let Assumption \ref{ass:cost} hold, and assume that the ED problem \eqref{eq:ED} is strictly feasible. There exists at least one competitive equilibrium. Furthermore, the production profile at any competitive equilibrium is efficient.
\end{propositio}
We omit the proof of Proposition \ref{thm:CE}, as it is straightforward to establish equivalence between the conditions for competitive equilibrium in Definition \ref{def:CE} and the (KKT) optimality  conditions for the original ED problem \eqref{eq:ED}.
It is important to mention that  the guaranteed efficiency of competitive equilibrium is a consequence of the particular nodal pricing mechanism that we employ---namely, \emph{locational marginal pricing}. We note that this is in contrast to the formulation of \cite{xiao2015efficiency}, which analyzes linear supply function equilibrium in a transmission constrained power network under a \emph{uniform pricing mechanism.} In the presence of transmissions constraints, uniform pricing mechanisms do not guarantee  the existence or the efficiency of a competitive equilibrium, in general.

\section{Example: Unbounded Price of Anarchy} \label{app:ex:unbounded_PoA}

Consider a two-node power network with a nodal demand profile given by $d = (D/2, D/2)$, where $D >0$.
Assume that the transmission line has capacity at least $c > D/2$, and that there are at least two producers at each node in the network.
Moreover,   producers that are common to a node are assumed to have identical production capacities. That is, we assume that $X_j = \capx_i$ for each producer $j \in \Ncal_i$ at each node $i \in \{1,2\}$. The production capacities are assumed to satisfy 
$$\frac{N_1 \capx_1}{D} \geq 1 > \frac{(N_1 - 1) \capx_1}{D} \quad \text{and} \quad \frac{(N_2 - 1) \capx_2}{D} > 1.$$
It follows that  Assumption \ref{ass:cap} is violated, as $\rsi_j < 1$ for  all producers $j \in \Ncal_1$. 
Define the production costs of producers according to
\begin{align*}
C_j(x_j) = \begin{cases}   \left(x_j\right)^+, &  \text{if} \   j \in \Ncal_1\\  \left(\beta (t)  x_j\right)^+, & \text{if} \ j \in \Ncal_2  \end{cases}
\end{align*}
where
\begin{align*}
\beta (t) =   \dfrac{ \left(1 + \frac{t / N_1}{(N_1 - 1) \capx_1 - t } \right)}{\left(1 + \frac{(D - t) / N_2}{(N_2 - 1) \capx_2 - (D - t) } \right)}
\end{align*}
for some parameter $t$ satisfying 
\begin{align*}
\frac{(N_1 - 1) \capx_1}{1+ \frac{N_2}{N_1} \left( \frac{(N_2 - 1) \capx_2}{D} - 1\right) }  < t < (N_1 - 1) \capx_1. 
\end{align*}
This guarantees that $\beta (t) > 1$. Given these assumptions, one can verify that the following strategy profile $(q, \theta)$ constitutes a Nash equilibrium for the game:
\begin{align*}
q &= (t, D - t) \\
\theta_j &= \left( 1 + \frac{t/N_1}{(N_1 - 1) \capx_1 - t} \right) \left(\capx_i - \frac{q_i}{ N_i}\right)
\end{align*}
for all $j \in \Ncal_i$ and  $i \in \{1, 2\}$. 
The production profile at the Nash equilibrium $(q, \theta)$ is therefore given by
\begin{align*}
x_j (q, \theta)= \begin{cases}
t / N_1, & \text{if } j \in \Ncal_1 \\
(D - t) / N_2, & \text{if } j \in \Ncal_2.
\end{cases}
\end{align*}
Also, the fact that $N_1 \capx_1 \geq D$  implies that an efficient production profile is given by
\begin{align*}
x_j^* = \begin{cases}
D / N_1, & \text{if } j \in \Ncal_1 \\
0, & \text{if } j \in \Ncal_2.
\end{cases}
\end{align*}
The price of anarchy associated with this game, therefore, satisfies
\begin{align*}
 \poa  &\geq \frac{  t + \beta(t) (D - t)}{ D},
\end{align*}
which implies that  $\poa \to \infty$ as $t \to (N_1 - 1) \capx_1$.

\section{Proof of Proposition \ref{prop:equilibrium}} \label{app:pf:prop:equilibrium}

The proof is divided into five parts. 
In part 1, we present necessary and sufficient optimality conditions for each producer's profit maximization problem and the ISO's economic dispatch (ED) problem. 
We use these conditions in parts 2 and 3 to  show that the production profile, $x(q, \theta)$, and nodal supply profile, $q$, at a Nash equilibrium $(q, \theta)$ are the unique optimal solutions to problem \eqref{opt:equilibrium} and \eqref{opt:equilibrium_q}, respectively.
In part 4, we show that the nodal price $p_i (q, \theta)$ at each node $i \in \Vcal$ at a Nash equilibrium $(q, \theta)$  satisfies conditions \eqref{eq:NEprice_1}--\eqref{eq:NEprice_2}. 
In part 5, we establish the existence of a Nash equilibrium by construction. 
Throughout the proof, we will assume that there are at least two producers at each node in the network, as this will serve to streamline the exposition. It is straightforward  to generalize the proof to accommodate scenarios in which  $N_i = 0$ for certain nodes $i \in \Vcal$ in the network.

\vspace{.1in}

\emph{Part 1 (Optimality Conditions):} \
Lemma \ref{lem:FoC_producer_j} provides a set of necessary and sufficient optimality conditions for each producer's profit maximization problem. Its proof is analogous to step 1 of the proof of \cite[Thm. 4.1]{Xu2014}, and is, therefore, omitted for the sake of brevity.
\begin{lemm} \label{lem:FoC_producer_j}
Let Assumptions \ref{ass:cost}-\ref{ass:cap} hold, and let  $q \in \Xcal_0 \cap \Rset_+^n$. For  each producer $j \in \Ncal_i$ at each node $i \in \Vcal$, 
$$\pi_j \left( q, \theta_j, \theta_{-j} \right) \geq \pi_j \left( q, \overline{\theta}_j, \theta_{-j} \right) \quad \forall \ \  \overline{\theta}_j \in \Xcal_j$$
if and only if the following conditions are satisfied:
\begin{align}
 x_j (q, \theta) \in [0, X_j ] ,\label{eq:optimality_j_pf_1}
\end{align}
and
\begin{align}
p_i (q, \theta) &\in  \left[ 0 , \ \frac{\partial^+ \widetilde{C}_j } { \partial x_j} \right] \hspace{3.55em} \text{if } \ x_j (q, \theta) = 0 ,\label{eq:optimality_j_pf_2}\\
p_i (q, \theta) & \in  \left[   \frac{\partial^- \widetilde{C}_j }{  \partial x_j} ,   \  \frac{ \partial^+ \widetilde{C}_j }{ \partial x_j}  \right]  \quad \ \text{if } \ x_j (q, \theta) \in (0, X_j) , \\
p_i (q, \theta) & \in  \left[ \frac{\partial^- \widetilde{C}_j }{  \partial x_j} , \ \infty \right)  \hspace{2.85em} \text{if } \ x_j (q, \theta) = X_j,
\label{eq:optimality_j_pf_3}
\end{align}
where the \emph{production quantity}  $x_j (q, \theta)$ is defined according to Eqs. \eqref{eq:production_1}-\eqref{eq:production_2},  and the \emph{nodal price} $p_i (q, \theta)$ is defined according to  Eq. \eqref{eq:price}. The left and right partial derivatives of the modified cost function are calculated at: 
\begin{align*}
\partial^- \widetilde{C}_j/\partial x_j &  :=   \ \partial^- \widetilde{C}_j (x_j (q, \theta); q_i) /\partial x_j,  \\
\partial^+ \widetilde{C}_j/\partial x_j & :=  \  \partial^+ \widetilde{C}_j (x_j (q, \theta); q_i) /\partial x_j.  
\end{align*}
\end{lemm}

\vspace{.1in}

Lemma  \ref{lem:FoC_ISO} provides  a set of necessary and sufficient optimality conditions for the ISO's ED problem \eqref{opt:mkt_clearing_q}.
\begin{lemm}\label{lem:FoC_ISO}
Let  $(q,\theta) \in \Xcal$, and assume that
\begin{align}
q_i < \sum\nolimits_{j \in \Ncal_i} X_j \label{eq:qi_condition}
\end{align}
for each node $i \in \Vcal$. Then, the nodal supply profile $q$ is the unique optimal solution to problem \eqref{opt:mkt_clearing_q} if and only if there exist multipliers $\lambda \in \Rset$ and $\mu \in \Rset_+^{2m}$ that satisfy 
\begin{align}
&p (q, \theta) = \lambda \bone  - H^\top \mu ,    \label{eq:KKT_0_pf_1} \\
&\mu \circ (H (q-d)- c) =0, \label{eq:KKT_0_pf_2}
\end{align}
where  the vector $p(q, \theta) := (p_1 (q, \theta), \dots, p_n (q, \theta))$ is determined according to Eq. \eqref{eq:price}. 
\end{lemm}

\begin{proof}[Proof of Lemma \ref{lem:FoC_ISO}]

Problem \eqref{opt:mkt_clearing_q} is a convex program with linear constraints. It follows that the Karush-Kuhn-Tucker (KKT) conditions are both necessary and sufficient for optimality. That is to say, $q \in \Xcal_0$ is optimal if and only if there exist Lagrange multipliers $\lambda \in \Rset$ and $\mu \in \Rset_+^{2m}$ associated with the constraints $\bone^\top (q - d) = 0$ and $H (q - d) \leq c$, respectively, which satisfy the \emph{stationarity conditions}
\begin{align}
\nabla  \left(  \sum_{i=1}^n  G_i(q_i; \theta) \right)   - \lambda \bone  + H^\top \mu  = 0,
\end{align}
and the \emph{complementary slackness conditions}
\begin{align*}
\mu \circ (H (q-d) - c) =0.
\end{align*}
Also, it is straightforward to show that 
\begin{align*}
G_i (q_i; \theta) =  \sum_{j \in \{\Ncal_i  |  \theta_j > 0\}}\theta_j \log \left( \frac{X_j \sum_{k \in \Ncal_i} \theta_k}{\theta_j \left( \sum_{k \in \Ncal_i} X_k - q_i \right)} \right) 
\end{align*}
if $\sum_{j \in \Ncal_i} \theta_j > 0$, and $G_i (q_i; \theta) = 0$ if $\sum_{j \in \Ncal_i} \theta_j = 0$. Hence, each function $G_i$ is differentiable in $q_i$ if inequality \eqref{eq:qi_condition} is satisfied. A direct calculation shows that 
\begin{align}
\nabla  \left(  \sum_{i=1}^n  G_i(q_i; \theta) \right)  = p(q, \theta),
\end{align}
where  the vector $p(q, \theta)$  is defined according to Eq. \eqref{eq:price}.
\end{proof}

\vspace{.1in}

\emph{Part 2 (Production Profile at Nash Equilibrium):} \ Let $(q, \theta)$ be a Nash equilibrium. We now prove that the production profile $x(q, \theta)$ (defined according to   Eqs. \eqref{eq:production_1}-\eqref{eq:production_2}) is the unique optimal solution to problem \eqref{opt:equilibrium}.
First,  it is straightforward to show that the modified cost function $\widetilde{C}_j (x_j; q_i)$ is strictly convex in $x_j$ over $[0, X_j]$,  given the satisfaction of  Assumptions \ref{ass:cost} and \ref{ass:cap}. This---in combination with fact that the feasible region of problem \eqref{opt:equilibrium}  is defined in terms of linear constraints---implies that the  KKT conditions \eqref{eq:primal_feas_1}-\eqref{eq:complementary} are necessary and sufficient for optimality. Specifically, a production profile $x \in \Rset^N$ is optimal for problem \eqref{opt:equilibrium} if and only if there exist  Lagrange multipliers $\lambda \in \Rset$ and $\mu \in \Rset_+^{2m}$ such that the following conditions hold.

\begin{enumerate}[(i)]\setlength{\itemsep}{.1in}
\item \emph{Primal feasibility:}
\begin{align}
&A x   - d \in \Pcal , \label{eq:primal_feas_1}\\
&x_j   \in [0,  X_j ] \quad \text{for } j = 1, \dots, N. \label{eq:primal_feas_2}
\end{align}
\item \emph{Stationarity:}
\end{enumerate}
\begin{align}
 p_i &\in \left(-\infty  , \ \frac{\partial^+ \widetilde{C}_j (x_j; q_i)}{ \partial x_j } \right] \hspace{4.85em} \text{if }  x_j = 0 , \label{eq:stationarity1}\\
 p_i & \in \left[ \frac{\partial^- \widetilde{C}_j (x_j; q_i) }{ \partial x_j}, \   \frac{\partial^+ \widetilde{C}_j  (x_j; q_i) }{ \partial x_j} \right] \quad  \text{if }  x_j   \in (0, X_j), \\
 p_i &\in \left[ \frac{\partial^- \widetilde{C}_j (x_j; q_i) }{ \partial x_j}, \ \infty \right) \hspace{5.7em} \text{if }  x_j  = X_j,
\label{eq:stationarity3}
\end{align}
\setlength{\leftskip}{2em}for each $i \in \Vcal$ and $j \in \Ncal_i$, where the vector $p \in \Rset^n$ is defined according to
\begin{align}
p := \lambda \bone  - H^\top \mu. \label{eq:stationarity4}
\end{align}

\setlength{\leftskip}{0em}

\begin{enumerate}
\item[(iii)]  \emph{Complementary slackness:}
\begin{align}
\mu \circ (H (A x  -d) - c) &=0. \label{eq:complementary}
\end{align}
\end{enumerate}

\begin{figure*}[b]
\hrulefill
\begin{align}
\Lambda_i(z) := \left\{  \lambda  \in \Rset \; \left| \; \exists x \in \Rset^N  \ \text{such that } \sum_{j \in \Ncal_i} x_j = z \ \ \text{and} \ \  x_j \in \argmin_{ \overline{x}_j \leq X_j} \left\{  \widetilde{C}_j (\overline{x}_j; z) - \lambda \overline{x}_j  \right\} \ \ \forall \  j \in \Ncal_i \;  \right. \right\} \label{eq:Lambda_i(z)_def}
\end{align}
\end{figure*}

We now employ Lemmas \ref{lem:FoC_producer_j} and \ref{lem:FoC_ISO} to  show that the KKT conditions  are satisfied  at $x = x(q, \theta)$.

\vspace{.05in}

\emph{Primal feasibility:}  Since $(q, \theta)$ is a Nash equilibrium, it follows from Lemma \ref{lem:FoC_producer_j} that  conditions \eqref{eq:optimality_j_pf_1}-\eqref{eq:optimality_j_pf_3} are satisfied for each $j \in \{1, \dots, N\}$.
The combination of condition \eqref{eq:optimality_j_pf_1} and the fact that $A x(q, \theta) = q \in \Xcal_0$ implies the satisfaction of the primal feasibility conditions \eqref{eq:primal_feas_1}-\eqref{eq:primal_feas_2}.

\vspace{.05in}

\emph{Stationarity:} Inequality \eqref{eq:optimality_j_pf_1} and Assumption \ref{ass:cap} together guarantee that
\begin{align}
0 \leq q_i \leq q_i^{\max} < \sum_{j \in \Ncal_i} X_j - \max_{k \in \Ncal_i} X_k \label{eq:qi_ub}
\end{align}
for each node $i \in \Vcal$. It follows that  $q_i < \sum_{j \in \Ncal_i} X_j$ for each node $i \in \Vcal$. And, since $(q, \theta)$ is a Nash equilibrium, it  follows from Lemma \ref{lem:FoC_ISO} that there exist multipliers $\lambda \in \Rset$ and $\mu \in \Rset_+^{2m}$ that satisfy Eqs. \eqref{eq:KKT_0_pf_1} and \eqref{eq:KKT_0_pf_2}. The satisfaction of Eq.  \eqref{eq:KKT_0_pf_1}, in combination with the conditions \eqref{eq:optimality_j_pf_2}--\eqref{eq:optimality_j_pf_3}, implies that $(x(q, \theta), \lambda, \mu)$ satisfy desired the set of stationarity conditions \eqref{eq:stationarity1}--\eqref{eq:stationarity4}.

\vspace{.05in}

\emph{Complementary slackness:} Since $(q, \mu)$ satisfy Eq. \eqref{eq:KKT_0_pf_2} and  $q = A x(q, \theta)$, it follows that 
\begin{align}
\mu \circ (H (A x(q,\theta)  -d) - c) =0.
\end{align}

\vspace{.1in}

\emph{Part 3 (Nodal Supply Profile at Nash Equilibrium):} \ Let $(q, \theta)$  be a Nash equilibrium. We now prove that the nodal supply profile $q$ is the unique optimal solution to problem \eqref{opt:equilibrium_q}. 
At the heart of our proof is the following technical lemma, which establishes the strict convexity of the function $\widetilde{G}_i$ over a superset of $[0, q_i^{\max}]$, and characterizes its left and right derivatives.
Its proof can be found in Appendix \ref{app:pf:lem:G_i}.

\begin{lemm} \label{lem:G_i}
Let Assumption \ref{ass:cost} hold. For each node $i \in \Vcal$, define the constant 
$$Q_i := \sum_{j \in \Ncal_i} X_j - \max_{k \in \Ncal_i} X_k,$$
and define the set $\Lambda_i (z)$ according to Eq. \eqref{eq:Lambda_i(z)_def}
for each  $z \in [0, Q_i)$.   
All of the following statements are true:
\begin{enumerate}[(i)]\setlength{\itemsep}{.1in}
\item The set $\Lambda_i (z)$ is compact for each $z \in [0 , Q_i )$.
\item The function $\widetilde{G}_i (\overline{q}_i) $ is strictly convex on $[0, Q_i )$.
\item The left and right derivatives of the function $\widetilde{G}_i (\overline{q}_i)$ satisfy
\begin{align*}
\begin{alignedat}{8}
\frac{\partial^- \widetilde{G}_i (z)}{\partial \overline{q}_i}  &= \inf   \Lambda_i (z) \quad && \forall  z \in (0,  Q_i ) , \\
\frac{\partial^+ \widetilde{G}_i (z)}{\partial \overline{q}_i}  &= \sup   \Lambda_i (z)  \quad && \forall z \in  [0,  Q_i ) .
\end{alignedat}
\end{align*}
\end{enumerate}
\end{lemm}

\vspace{.1in}

Assumption \ref{ass:cap} implies that any feasible solution $\overline{q} \in \Xcal_0 \cap \Rset_+^n$ to problem \eqref{opt:equilibrium_q} is guaranteed to satisfy 
\begin{align}
0 \leq \overline{q}_i \leq q_i^{\max} < Q_i \label{eq:feasible_overline_q}
\end{align}
for each node $i \in \Vcal$. This implies that the objective function of problem \eqref{opt:equilibrium_q} is \emph{strictly convex} over its feasible region, which is defined by linear constraints. Therefore, $q$ is the unique optimal solution to problem \eqref{opt:equilibrium_q} if and only if there exist Lagrange multipliers $\lambda \in \Rset$ and $\mu \in \Rset_+^{2m}$ such that $(q, \lambda, \mu)$ satisfy the following KKT conditions.

\vspace{.1in}

\begin{enumerate}[(i)]\setlength{\itemsep}{.1in}
\item  \emph{Primal feasibility:} $q \in \Xcal_0 \cap \Rset_+^n$.
\item \emph{Stationarity:}
\begin{align}
p_i & \in \left( -\infty,  \  \frac{\partial^+ \widetilde{G}_i (q_i)}{\partial \overline{q}_i} \right] \hspace{3.35em}  \text{if } \ q_i = 0 , \label{eq:stationarity_q1}\\
p_i & \in \left[ \frac{\partial^- \widetilde{G}_i (q_i)}{\partial \overline{q}_i} , \  \frac{\partial^+ \widetilde{G}_i (q_i)}{\partial \overline{q}_i} \right] \quad \text{if } \ q_i > 0 , \label{eq:stationarity_q2}
\end{align}
for each node $i \in \Vcal$, where
\begin{align}
p = \lambda \bone - H^\top \mu. \label{eq:stationarity_q3}
\end{align}
\item   \emph{Complementary slackness:}
\begin{align}
\mu \circ (H (q-d)-c) &=0. \label{eq:complementary_q}
\end{align}
\end{enumerate}

The assumption that $(q, \theta)$ is a Nash equilibrium implies that  $q \in \Xcal_0 \cap \Rset_+^n$ (cf. the inequality in  \eqref{eq:qi_ub}).
We now  establish the existence of Lagrange multipliers $\lambda \in \Rset$ and $\mu \in \Rset_+^{2m}$ such that the stationarity and complementary slackness conditions are satisfied at
\begin{align}
p_i (q, \theta) & \in \left( -\infty,  \  \frac{\partial^+ \widetilde{G}_i (q_i)}{\partial \overline{q}_i} \right] \hspace{3.35em}  \text{if } \ q_i = 0 , \label{eq:KKT_q_pf1}\\
p_i (q, \theta) & \in \left[ \frac{\partial^- \widetilde{G}_i (q_i)}{\partial \overline{q}_i} , \  \frac{\partial^+ \widetilde{G}_i (q_i)}{\partial \overline{q}_i} \right] \quad \text{if } \ q_i > 0  ,  \label{eq:KKT_q_pf2}
\end{align}
for each node $i \in \Vcal$, and 
\begin{align}
p(q, \theta)  = \lambda \bone - H^\top \mu,   \label{eq:eyb1} \\
\mu \circ (H (q-d)-c) =0.  \label{eq:eyb2} 
\end{align}
Given the assumption  that $(q, \theta)$ is a Nash equilibrium, it follows from Lemma \ref{lem:FoC_producer_j}  that conditions \eqref{eq:optimality_j_pf_2}--\eqref{eq:optimality_j_pf_3} are satisfied for each  $j \in \Ncal_i$  and $i \in \Vcal$.
It is straightforward to verify that this implies that $p_i (q, \theta) \in \Lambda_i (q_i)$ for all $i \in \Vcal$.
It  follows from Lemma \ref{lem:G_i} that  conditions \eqref{eq:KKT_q_pf1} and \eqref{eq:KKT_q_pf2} are both satisfied. Finally, Lemma \ref{lem:FoC_ISO} guarantees the existence Lagrange multipliers $\lambda \in \Rset$ and $\mu \in \Rset_+^{2m}$  such that Eqs.  \eqref{eq:eyb1}--\eqref{eq:eyb2} are satisfied.

\vspace{.1in}

\emph{Part 4: (Nodal Price at Nash Equilibrium): } \ Let $(q, \theta)$ be a Nash equilibrium. 
The fact that the nodal price $p_i (q, \theta)$ at each node $i \in \Vcal$ satisfies conditions \eqref{eq:NEprice_1}--\eqref{eq:NEprice_2} immediately follows from conditions \eqref{eq:optimality_j_pf_2}--\eqref{eq:optimality_j_pf_3} in Lemma \ref{lem:FoC_producer_j}, as the left derivative of the modified cost function $\widetilde{C}_j (x_j; q_i)$ evaluated at $x_j = 0$ satisfies $\partial^- \widetilde{C}_j (0; q_i) / \partial x_j = 0$.

\vspace{.1in}

\emph{Part 5 (Construction of a Nash Equilibrium):} \ Before proceeding with the proof, we first provide a sketch of the main arguments employed. We establish the existence of a Nash equilibrium by construction. 
 We do so, in part 5-A, by first constructing  a nodal supply profile $q \in \Rset_+^n$ as the unique optimal solution to problem \eqref{opt:equilibrium_q}. 
Next, we solve an equivalent reformulation of problem \eqref{opt:equilibrium} to obtain a production profile $x \in \Rset_+^N$, and define a nodal price vector $p \in \Rset_+^n$  according to a linear combination of its optimal Lagrange multipliers.
Using the resulting production quantities and nodal prices, we construct  a producer strategy profile according to 
\begin{align*}
\theta_j := p_i(X_j - x_j)
\end{align*}
for each producer $j \in \Ncal_i$ and node $i \in \Vcal$. 
In part 5-B, we complete the proof by showing that the necessary and sufficient conditions for Nash equilibrium established in Lemmas \ref{lem:FoC_producer_j} and \ref{lem:FoC_ISO}  are indeed satisfied by the pair $(q, \theta)$ as constructed.

\vspace{.1in}

\emph{Part 5-A (Constructing a Candidate NE):} \ 
Let $q$ be the unique optimal solution to problem \eqref{opt:equilibrium_q}. Consider the following relaxation to problem \eqref{opt:equilibrium}, where we have dropped the nonnegativity constraints on the production quantities.
\begin{equation}
\begin{alignedat}{8}
&\underset{\overline{x} \in \Rset^N}{\text{minimize}} \quad & &\sum_{i=1}^n \sum_{j \in \Ncal_i} \widetilde{C}_j (\overline{x}_j; q_i  )\\
&\text{subject to} \quad & & A\overline{x} - d \in \Pcal, \\
&&&  \overline{x}_j \leq X_j, \quad   j = 1, \dots, N.
\end{alignedat} \label{opt:equilibrium_relaxed}
\end{equation}
Assumptions \ref{ass:cost} and \ref{ass:cap} guarantee that each modified cost function $\widetilde{C}_j (\overline{x}_j; q_i) $ is strictly convex and strictly increasing in $\overline{x}_j$ over $[0, X_j]$, and is equal to zero  for all values $\overline{x}_j \leq 0$. This guarantees that problem \eqref{opt:equilibrium_relaxed} has a unique optimal solution that is nonnegative elementwise, which, in turn, implies the optimality of this solution for problem \eqref{opt:equilibrium}.

Let $x \in \Rset^N_+$ be the unique optimal solution to problem \eqref{opt:equilibrium_relaxed}, which is a convex program with linear constraints. It follows that there exist Lagrange multipliers $\lambda \in \Rset$ and $\mu \in \Rset_+^{2m}$ such that the following conditions are satisfied.

\vspace{.05in}

\begin{enumerate}[(i)]
\item \emph{Stationarity:} For each $i \in \Vcal$ and $j \in \Ncal_i$,
\begin{align}
&p_i \in \left[ \frac{\partial^- \widetilde{C}_j (x_j; q_i)}{\partial \overline{x}_j} ,  \frac{\partial^+ \widetilde{C}_j (x_j; q_i)}{\partial \overline{x}_j} \right] && \text{if } x_j  < X_j \label{eq:stationarity_relaxed1} \\
&p_i \in \left[ \frac{\partial^- \widetilde{C}_j (x_j; q_i)}{\partial \overline{x}_j} , \  + \infty \right)  && \text{if } x_j  = X_j \label{eq:stationarity_relaxed2}
\end{align}
 where 
\begin{align}
p := \bone \lambda - H^\top \mu. \label{eq:stationarity_relaxed3}
\end{align}
\item \emph{Complementary slackness:}
\begin{align}
\mu \circ (H (q-d)-c) =0. \label{eq:complementary_relaxed}
\end{align}
\end{enumerate}
The vector $p$ that we specify in Eq. \eqref{eq:stationarity_relaxed3} will  play the role of  a `nodal price vector' in constructing producers' strategy profile. 
Specifically, we construct the producers' strategy profile $\theta$ according to
\begin{align*}
\theta_j := p_i (X_j - x_j )  
\end{align*}
for each  producer $j \in \Ncal_i$ and node $i \in \Vcal$.  As constructed, the producers' strategy profile $\theta$ is guaranteed to be nonnegative elementwise, as the inequalities \eqref{eq:stationarity_relaxed1}--\eqref{eq:stationarity_relaxed2} guarantee that $p_i \geq 0$ for each $i \in \Vcal$, since each modified function $\widetilde{C}_j$ is non-decreasing over $(-\infty, X_j]$.

\vspace{.1in}

\emph{Part 5-B (Checking Necessary and Sufficient NE Conditions):} \ Recall that $x \in \Rset_+^N$ and $q \in \Rset_+^n$ are the unique optimal solutions to problems \eqref{opt:equilibrium_relaxed} and \eqref{opt:equilibrium_q}, respectively. It is not difficult to show that $x$ and $q$ are  related according to
\begin{align*} 
Ax = q. 
\end{align*}
Using this fact, it is straightforward to verify that the nodal prices  $p(q, \theta)$ and production quantities $x(q,\theta)$ induced by $(q, \theta)$ satisfy
\begin{align}
p_i (q, \theta) &= p_i  \quad \forall \ i \in \Vcal, \label{eq:priceNash} \\
x_j (q, \theta) &= x_j  \quad \forall \  j \in \{1, \dots, N\},  \label{eq:xNash}
\end{align}
 where  recall that $p_i (q, \theta)$  is defined according to Eq. \eqref{eq:price}, and $x_j (q, \theta)$  is defined according to Eqs. \eqref{eq:production_1}--\eqref{eq:production_2}. 
 
 To complete the proof, it suffices to show that the necessary and sufficient conditions for Nash equilibrium established in Lemmas \ref{lem:FoC_producer_j} and \ref{lem:FoC_ISO}  are satisfied by the pair $(q, \theta)$ as constructed. This is immediate to see upon examination of Eqs.  \eqref{eq:stationarity_relaxed1}--\eqref{eq:xNash}.
 This completes the proof that $(q, \theta)$ is a Nash equilibrium.


\section{Proof of Lemma \ref{lem:G_i}} \label{app:pf:lem:G_i}

\emph{Proof of statement (i):} \ Fix $i \in \Vcal$ and  $ z \in [0, Q_i)$. Consider the following convex optimization problem:
\begin{equation}
\begin{alignedat}{8}
&\underset{x \in \Rset^N}{\text{minimize}} \quad &&\sum_{j \in \Ncal_i} \widetilde{C}_j (x_j; z) \\
&\text{subject to} \quad && \sum_{j \in \Ncal_i} x_j = z \\
&&&  x_j \leq X_j, \quad && \text{if } \   j \in \Ncal_i \\
&&& x_j = 0, \quad && \text{if } \  j \in \{1, \dots, N\} \setminus \Ncal_i,
\end{alignedat} \label{opt:node_i}
\end{equation}
It follows from Assumption \ref{ass:cost}  that the modified cost function $\widetilde{C}_j$ is strictly convex and strictly increasing over $[0, X_j]$, and satisfies $\widetilde{C}_j (x_j; z) = 0$ for $x_j \leq 0$.
This guarantees that problem \eqref{opt:node_i} admits a unique optimal solution $\widetilde{x}(z)$ that is non-negative element-wise. 
Notice that the set $\Lambda_i (z)$  is the set of optimal Lagrange multipliers associated with the nodal power balance constraint $\sum_{j \in \Ncal_i} x_j = z$ for problem \eqref{opt:node_i}.
It can, therefore, be expressed as
\begin{align*}
\Lambda_i (z) = \Bigg\{ \lambda \; \Bigg| \; \frac{\partial^- \widetilde{C}_j (\widetilde{x}_j (z); z)}{\partial x_j}  \leq \lambda \leq  \frac{\partial^+ \widetilde{C}_j (\widetilde{x}_j (z); z)}{\partial x_j},& \\ 
\forall j \in \Ncal_i &\Bigg\}.
\end{align*}
The above expression can be simplified to
\begin{align}
\Lambda_i (z) = \left[ \max_{j \in \Ncal_i} \left\{ \frac{\partial^- \widetilde{C}_j (\widetilde{x}_j (z); z)}{\partial x_j}  \right\}, \, \min_{j \in \Ncal_i} \left\{ \frac{\partial^+ \widetilde{C}_j (\widetilde{x}_j (z); z)}{\partial x_j}  \right\} \right]. \label{eq:Lambda_i(z)}
\end{align}
It follows that set $\Lambda_i (z)$ is compact for every $z \in [0, Q_i)$.

\vspace{.1in}

\emph{Proof of statement (ii):} \ 
It suffices to show that $g_i (z)$ is strictly increasing in $z$ over $(0, Q_i)$. 
Fix $0 < z < y < Q_i$. We  now show that $g_i (z) < g_i (y)$.
First note that $g_i (z)$ satisfies
\begin{align}
g_i (z) = \max_{j \in \Ncal_i} \left\{ \frac{\partial^- \widetilde{C}_j (\widetilde{x}_j (z); z)}{\partial x_j}  \right\} = \inf \Lambda_i (z), \label{eq:g_i(z)}
\end{align}
where 
\begin{align}
\frac{\partial^- \widetilde{C}_j (\widetilde{x}_j (z); z)}{\partial x_j}  = \frac{\partial^- C_j (\widetilde{x}_j (z))}{\partial x_j}  \left( 1 + \frac{\widetilde{x}_j (z)}{\sum_{k \in \Ncal_i, k \neq j} X_k - z  } \right). \label{eq:derivative_tilde_C}
\end{align}
for each $j \in \Ncal_i$.
Additionally, $0 < z < y < Q_i$ implies that there exists $j_0 \in \Ncal_i$, such that
$$0 \leq \widetilde{x}_{j_0} (z) < \widetilde{x}_{j_0} (y) \leq X_j.$$
This implies the following chain of inequalities:
\begin{align*}
g_i (z) & = \inf \Lambda_i (z) \leq \sup \Lambda_i (z) \leq \frac{\partial^+ \widetilde{C}_{j_0} (\widetilde{x}_{j_0} (z); z)}{\partial x_{j_0}}  \\
& < \frac{\partial^- \widetilde{C}_{j_0} (\widetilde{x}_{j_0} (y); z)}{\partial x_{j_0}} \\
& = \frac{\partial^- C_{j_0} (\widetilde{x}_{j_0} (y))}{\partial x_{j_0}}  \left( 1 + \frac{\widetilde{x}_{j_0} (y)}{\sum_{k \in \Ncal_i, k \neq j_0} X_k - z  } \right) \\
& < \frac{\partial^- C_{j_0} (\widetilde{x}_{j_0} (y))}{\partial x_{j_0}}  \left( 1 + \frac{\widetilde{x}_{j_0} (y)}{\sum_{k \in \Ncal_i, k \neq j_0} X_k - y  } \right) \\
& = \frac{\partial^- \widetilde{C}_{j_0} (\widetilde{x}_{j_0} (y); y)}{\partial x_{j_0}} \\ &\leq g_i (y).
\end{align*}
Here, the first line follows from a combination of Eqs. \eqref{eq:Lambda_i(z)} and \eqref{eq:g_i(z)}; the second line follows from the strict convexity of the function $\widetilde{C}_j (x_j; z)$ in $x_j$ over $[0, X_j]$; the third and the fifth lines follow from Eq. \eqref{eq:derivative_tilde_C}; and the fourth line follows from the fact that $0 < z < y < Q_i$. This finishes the proof of statement (ii).

\vspace{.1in}

\emph{Proof of statement (iii):} \ It will be convenient to define the function
\begin{align*}
h_i (z) := \sup \Lambda_i (z).
\end{align*}
Using the fact that convex functions defined over a bounded interval are differentiable at all but countably many points in that interval, it is straightforward to show that $h_i (z) = g_i (z)$ for all but countably many $z \in [0, Q_i )$.
This  implies that
\begin{align}
\widetilde{G}_i (\overline{q}_i) = \int_{0}^{\overline{q}_i} g_i (z) \mathrm{d} z = \int_{0}^{\overline{q}_i} h_i (z) \mathrm{d} z. \label{eq:tilde_G_i}
\end{align}
Consequently, statement (iii) is true if $h_i (z)$ is right-continuous on $[0, Q_i)$ and $g_i (z)$  is left-continuous on $(0, Q_i)$.
We only prove the left-continuity of the function $g_i (z)$, as the proof of right-continuity of $h_i (z)$ is analogous.

First, it follows from  Eq. \eqref{eq:derivative_tilde_C} and the convexity of $\widetilde{C}_j$  that $g_i (z)$ is left-continuous if $\widetilde{x}_j (z)$ is left-continuous in $z$ for each $j \in \Ncal_i$. 
Recall that for any $z \in (0, Q_i )$,  the parametric optimizer $\widetilde{x} (z)$ is the unique optimal solution to problem \eqref{opt:node_i}, and is guaranteed to be nonnegative elementwise. This implies that $\widetilde{x} (z)$ is also the unique optimal solution of the following convex program:
\begin{equation}
\begin{alignedat}{8}
&\underset{x \in \Rset^N}{\text{minimize}} \quad &&\sum_{j \in \Ncal_i} \widetilde{C}_j (x_j; z) \\
&\text{subject to} \quad && \sum_{j \in \Ncal_i} x_j = z \\
&&&  0 \leq x_j \leq X_j, \quad && \text{if } \   j \in \Ncal_i \\
&&& x_j = 0, \quad && \text{if } \  j \in \{1, \dots, N\} \setminus \Ncal_i.
\end{alignedat} \label{opt:node_i_equivalent}
\end{equation}
The feasible region of problem \eqref{opt:node_i_equivalent} is compact and continuous (i.e., both upper and lower hemicontinuous) in the parameter $z$ over $(0, Q_i)$. Additionally, the objective function of problem \eqref{opt:node_i_equivalent} is strictly convex in $x$ and jointly continuous in $(x,z)$ for all $z \in (0, Q_i)$. 
It follows from Berge's maximum theorem in \cite[p. 116]{berge1963topological} that the unique parametric optimizer $\widetilde{x} (z)$ of problem \eqref{opt:node_i_equivalent} is continuous in $z$ on $(0, Q_i)$. In particular, this implies the left-continuity of $\widetilde{x}_j (z) $, which completes the proof.

\section{Proof of Lemma \ref{lem:NE_2node}} \label{app:pf:lem:NE_2node}

The crux of the proof centers on the derivation of a closed-form expression for the optimal solution of problem \eqref{opt:2node}. We first eliminate the decision variable $\overline{q}_2$  through   substitution of the power balance constraint $\overline{q}_2 = D - \overline{q}_1$, which yields the   equivalent reformulation of  problem \eqref{opt:2node} as:
\begin{equation}
\begin{alignedat}{8}
& \underset{\overline{q}_1 \in \Rset}{\text{minimize}} \quad && \quad \int_0^{\overline{q}_1} \beta_1 \left( 1 +  \frac{1/N_1}{(N_1 - 1) \capx_1 / z - 1} \right) \mathrm{d} z \\
&&& + \int_0^{D - \overline{q}_1} \beta_2 \left( 1 +  \frac{1/N_2}{(N_2 - 1) \capx_2 / z - 1} \right) \mathrm{d} z \\
& \text{subject to} \quad && (d_1 - c)^+ \leq \overline{q}_1 \leq D - (d_2 - c)^+.
\end{alignedat} \label{opt:2node_univariate}
\end{equation}
It will be notationally convenient to denote the projection operator onto a closed interval $[a,b]  \subseteq \Rset$ according to $[\cdot]_{a}^b$.
It is straightforward to show that the unique optimal solution to problem \eqref{opt:2node_univariate} is given by
\begin{align}
q_1  =  \left[ \widetilde{q}_1 \right]_{d_1 - c}^{D - (d_2 - c)^+}, \label{eq:q1_opt_2node}
\end{align}
where $\widetilde{q}_1$ is the unique solution to the following first-order condition on the open interval $(0, (N_1 - 1) \capx_1)$:
\begin{equation}
\begin{split}
\beta_1\left( 1+  \left( N_1 \left( \frac{\capx_1 (N_1 - 1)}{\overline{q}_1} - 1 \right)\right)^{-1} \right)  \\
= \beta_2\left( 1 + \left( N_2 \left( \frac{\capx_2 (N_2 - 1)}{D - \overline{q}_1} - 1 \right)\right)^{-1} \right).
\end{split} \label{eq:q1_relax}
\end{equation}
 It follows from Eq. \eqref{eq:q1_opt_2node} that
\begin{align*}
\cost_{\nash} = \beta_2 D + (\beta_1 - \beta_2) \left[ \widetilde{q}_1 \right]_{d_1 - c}^{D - (d_2 - c)^+}.
\end{align*}
Since $\widetilde{q}_1$ does not depend on the transmission capacity $c$, it holds that
\begin{align*}
\frac{\partial^+ \cost_{\nash}}{\partial c} = \begin{cases}
\beta_2 - \beta_1 & \text{if } \widetilde{q}_1 < d_1 - c ,\\
\beta_1 - \beta_2 & \text{if } \widetilde{q}_1 > D - (d_2 - c)^+  \ \text{and } c < d_2,\\
0 & \text{otherwise}.
\end{cases}
\end{align*}
To complete the proof of the first part of Lemma \ref{lem:NE_2node}, it suffices to show that the following two conditions hold:
\begin{align}
 \label{cond1} p_1 > p_2 & \iff \widetilde{q}_1 < d_1 - c, \\
\label{cond2} p_1 < p_2 & \iff \widetilde{q}_1 > D - (d_2 - c)^+  \ \text{and } c < d_2.
\end{align}
To show that condition \eqref{cond1} holds, first recall the explicit formulae for the LMPs at Nash equilibrium in  Eqs. \eqref{eq:p_2node_case1}-\eqref{eq:p_2node_case2}. It follows that  $p_1 > p_2$ if and only if
\begin{equation}
\begin{split}
&\beta_1\left( 1 + \left( N_1 \left( \frac{\capx_1 (N_1 - 1)}{q_1} - 1 \right)\right)^{-1} \right)  \\
> &\beta_2 \left( 1 +  \left( N_2 \left( \frac{\capx_2 (N_2 - 1)}{D - q_1} - 1 \right)\right)^{-1} \right),
\end{split} \label{eq:pf_p1>p2}
\end{equation}
where  $q_1$ is  specified according to Eq. \eqref{eq:q1_opt_2node}. It follows that 
\begin{align*}
p_1 > p_2  \iff   \widetilde{q}_1 < q_1,
\end{align*}
since $\widetilde{q}_1$ is the unique solution to  Eq. \eqref{eq:q1_relax}. Furthermore, it holds that 
\begin{align*}
\widetilde{q}_1 < q_1 \iff \widetilde{q}_1 < d_1 - c,
\end{align*}
since $q_1  =  \left[ \widetilde{q}_1 \right]_{d_1 - c}^{D - (d_2 - c)^+}$.  This proves that \eqref{cond1} holds. The proof that  \eqref{cond2} holds is analogous, and is, therefore, omitted.

To prove the second part of Lemma \ref{lem:NE_2node}, note that the previous arguments also imply that $p_1 > p_2$ if and only if the inequality in \eqref{eq:pf_p1>p2} is satisfied for $q_1 = d_1 - c$. Plugging $q_1 = d_1 - c$ into \eqref{eq:pf_p1>p2} shows that $p_1 > p_2$ if and only if the inequality in \eqref{eq:braess} holds.

\end{appendices}

\end{document}